\documentclass[useAMS,usenatbib]{mn2e}

\usepackage{natbib}
\usepackage{amsmath}
\usepackage[dvips]{graphicx}
\usepackage{longtable}
\usepackage{float}
\usepackage{caption}
\usepackage{subfig}
\usepackage{txfonts}
\usepackage{epsfig}
\usepackage{dcolumn}
\newcolumntype{d}{D{.}{.}{-1}}
\def\dhead#1{\multicolumn{1}{c}{#1}}
\def\ehead#1{\multicolumn{1}{l}{#1}}
\graphicspath{{.}{figs/}}
\newenvironment{noindlist}
 {\begin{list}{\labelitemi}{\leftmargin=0em \itemindent=0em}}
 {\end{list}}
\title[10C Survey -- First Results]
  {10C Survey of Radio Sources at 15.7~GHz: II -- First Results\thanks{We
   request that any reference to this paper cites `AMI Consortium: Davies et al. 2010'}}
\author[AMI Consortium: Davies et al.]
  {AMI Consortium:
  Matthew~L.~Davies,$^1$\thanks{Email: m.davies@mrao.cam.ac.uk}  
  Thomas~M.~O.~Franzen,$^1$\thanks{Email: t.franzen@mrao.cam.ac.uk}
  \newauthor
  Elizabeth~M.~Waldram,$^1$
  Keith~J.~B.~Grainge,$^{1,2}$
  Michael~P.~Hobson,$^1$  
  \newauthor
  Natasha~Hurley-Walker,$^1$  
  Anthony~Lasenby,$^{1,2}$
  Malak~Olamaie,$^1$
  Guy~G.~Pooley,$^1$  
  \newauthor
  Julia~M.~Riley,$^1$
  Carmen Rodr\'{i}guez-Gonz\'{a}lvez,$^1$  
  Richard~D.~E.~Saunders,$^{1,2}$
  \newauthor  
  Anna~M.~M.~Scaife,$^3$  
  Michel~P.~Schammel,$^1$
  Paul~F.~Scott,$^1$
  Timothy~W.~Shimwell,$^1$
  \newauthor
  David~J.~Titterington,$^1$  
  and Jonathan~T.~L.~Zwart$^4$\\
  $^1$Astrophysics Group, Cavendish Laboratory,
      19 J.~J.~Thomson Avenue, Cambridge CB3 0HE \\
  $^2$Kavli Institute for Cosmology Cambridge,
      Madingley Road, Cambridge, CB3 0HA \\
  $^3$Dublin Institute for Advanced Studies,
      31 Fitzwilliam Place, Dublin 2, Ireland \\
  $^4$Columbia Astrophysics Laboratory, Columbia
      University, 550 West 120th Street, New York,
      NY 10027, U.S.A.}
\date{Accepted ????. Received ????}
\pagerange{\pageref{firstpage}--\pageref{lastpage}}
\pubyear{2010}

\begin{document}
\maketitle
\label{firstpage}

\begin{abstract}

\noindent
In a previous paper, the observational, mapping and
source-extraction techniques used for the Tenth Cambridge (10C)
Survey of Radio Sources were described.  Here, the first results
from the survey, carried out using the Arcminute Microkelvin Imager
Large Array (LA) at an observing frequency of 15.7~GHz, are
presented.  The survey fields cover an area of
$\approx 27$~deg$^{2}$ to a flux-density completeness of 1~mJy.
Results for some deeper areas, covering $\approx 12$~deg$^{2}$,
wholly contained within the total areas and complete to 0.5~mJy,
are also presented.  The completeness for both areas is estimated
to be at least 93~per~cent.  The 10C survey is the deepest radio
survey of any significant extent ($\gtrsim 0.2$~deg$^{2}$) above
1.4~GHz.

The 10C source catalogue contains 1897 entries and is available
online from the survey website (www.mrao.cam.ac.uk/surveys/10C).
The source catalogue has been combined with that of the Ninth
Cambridge Survey to calculate the 15.7-GHz source counts.  A broken
power law is found to provide a good parameterisation of the
differential count between 0.5~mJy and 1~Jy.  The measured source
count has been compared to that predicted by \citet{dezotti2005}
-- the model is found to display good agreement with the data at
the highest flux densities.  However, over the entire flux-density
range of the measured count (0.5~mJy to 1~Jy), the model is found
to under-predict the integrated count by $\approx 30$ per cent.

Entries from the source catalogue have been matched to those
contained in the catalogues of the NRAO-VLA Sky Survey and the
Faint Images of the Sky at Twenty Centimetres survey (both of which
have observing frequencies of 1.4~GHz).  This matching provides
evidence for a shift in the typical 1.4-to-15.7-GHz spectral index
of the 15.7-GHz-selected source population with decreasing flux
density towards sub-mJy levels -- the spectra tend to become less
steep.

Automated methods for detecting extended sources, developed in the
earlier 10C paper, have been applied to the data;
$\approx 5$~per~cent of the sources are found to be extended
relative to the LA synthesised beam of $\approx 30$~arcsec.
Investigations using higher-resolution data showed that most of
the genuinely extended sources at 15.7~GHz are classical doubles,
although some nearby galaxies and twin-jet sources were also
identified.

\end{abstract}

\nokeywords

\section{Introduction}\label{Introduction}

\subsection{Background}

The Ninth Cambridge (9C) Survey of Radio Sources 
\citep{waldram2003,waldram2010}, carried out using the Ryle
Telescope (RT) at an observing frequency of 15.2 GHz, was a
milestone in the exploration of the high-radio-frequency sky,
as the first survey of significant extent and depth at
such a high radio frequency.  Since the publication of the first 9C
paper, extensive survey work has been carried out using the
Australia Telescope Compact Array at 20~GHz
\citep[ATCA;][]{ricci2004,sadler2006,massardi2008,massardi2010,
murphy2010}.  The two surveys are complementary, with 9C probing
deeper flux-density levels (down to 5.5~mJy) and the ATCA surveys
covering shallower and wider areas (most recently, the whole
southern sky).

It is well known that high-frequency radio surveys are highly
time-consuming.  The scaling of interferometer primary-beam
areas with frequency ($\propto \nu^{-2}$), and the
typical synchrotron spectra of radio sources ($\propto \nu^{-0.7}$)
conspire so that the time required to carry out a survey of
equivalent depth and sky-coverage, using a telescope of fixed
aperture diameter, scales as $\nu^{3.4}$.  Things are somewhat better
if it is assumed that the available bandwidth scales linearly with
frequency.  However, the fact that the noise temperatures of the
available front-end, low-noise amplifiers used in interferometers
tends to increase with frequency must also be taken into account.

For these reasons, relatively little survey work has been undertaken
at high radio frequencies and the knowledge of the source population
remains poor.  Nevertheless, familiarity with the properties of this
population is important for the interpretation of the results from 
observations of the Cosmic Microwave Background (CMB), such as those
made by \textit{Planck} \citep{tauber2010}.  At mm wavelengths, foreground
radio sources are the dominant source of contamination of small-scale
CMB anisotropies \citep{dezotti1999}.  \citet{waldram2003,waldram2010}
have demonstrated that extrapolation of the flux densities of sources
at low frequencies cannot be relied upon to predict their high-frequency
properties, which emphasises the value of survey work at the frequencies
of interest ($\gtrsim 10$~GHz) for CMB work.

Samples of bright sources selected at high radio frequencies have
significant proportions with flat or rising spectra 
\citep[see, for example,][]{taylor2001,davies2009}.  In the main, these
sources are believed to be blazars with synchrotron self-absorbed spectra;
the self-absorbed components of such sources are often highly variable
\citep[see, for example,][]{franzen2009}.

High-frequency-selected samples also include appreciable numbers of
sources with convex spectra, peaking at GHz frequencies
\citep[see, for example,][]{bolton2004}.  Some of these GHz peaked
spectrum (GPS) sources \citep[see][for a review]{odea1998} are believed
to be associated with young objects, which later expand into powerful
radio sources, though many are dominated by emission from a strongly-beamed
self-absorbed component \citep{bolton2006}.  Surveys such as
the 9C provide flux-density-limited samples, which are useful for
gaining further understanding of the evolution of such objects.

\subsection{This work}

Since the 9C survey was carried out, the RT has
been transformed, by the installation of new front-end receivers
and back-end electronics (including a new correlator), into the
Arcminute Microkelvin Imager (AMI) Large Array (LA)
\citep[see][for a detailed description of the telescope]{zwart2008}.
The LA is a radio synthesis-telescope, located $\approx 19$~m above
sea level near Cambridge.  It is used to observe at a centre frequency
of 15.7~GHz and has a usable bandwidth of 4.5~GHz.  At this frequency,
the telescope has a full-width-at-half-maximum (FWHM) primary beam
of $\approx 5.5$~arcmin and a resolution of $\approx 30$~arcsec.

The LA has been used to carry out the Tenth Cambridge (10C) Survey
of Radio Sources.  As part of this survey, the improved flux
sensitivity of the LA, compared with the RT, has been used to explore
the 15-GHz-band sky to sub-mJy levels.  In a previous paper
\citep[hereafter Paper~I]{franzen2010} detailed technical information
regarding the survey strategy, mapping and source-extraction
techniques for the 10C survey was provided.  In this paper, the first
results from 10 fields, including the 15.7-GHz source count, are
presented.  Throughout this paper any equatorial coordinates use 
equinox J2000 and spectral indices are defined using the convention
that $S \propto \nu^{-\alpha}$.

\section{The 10C source catalogue}\label{The 10C source catalogue}

\begin{figure*}
 \centerline{\includegraphics[width=13cm, angle=270]{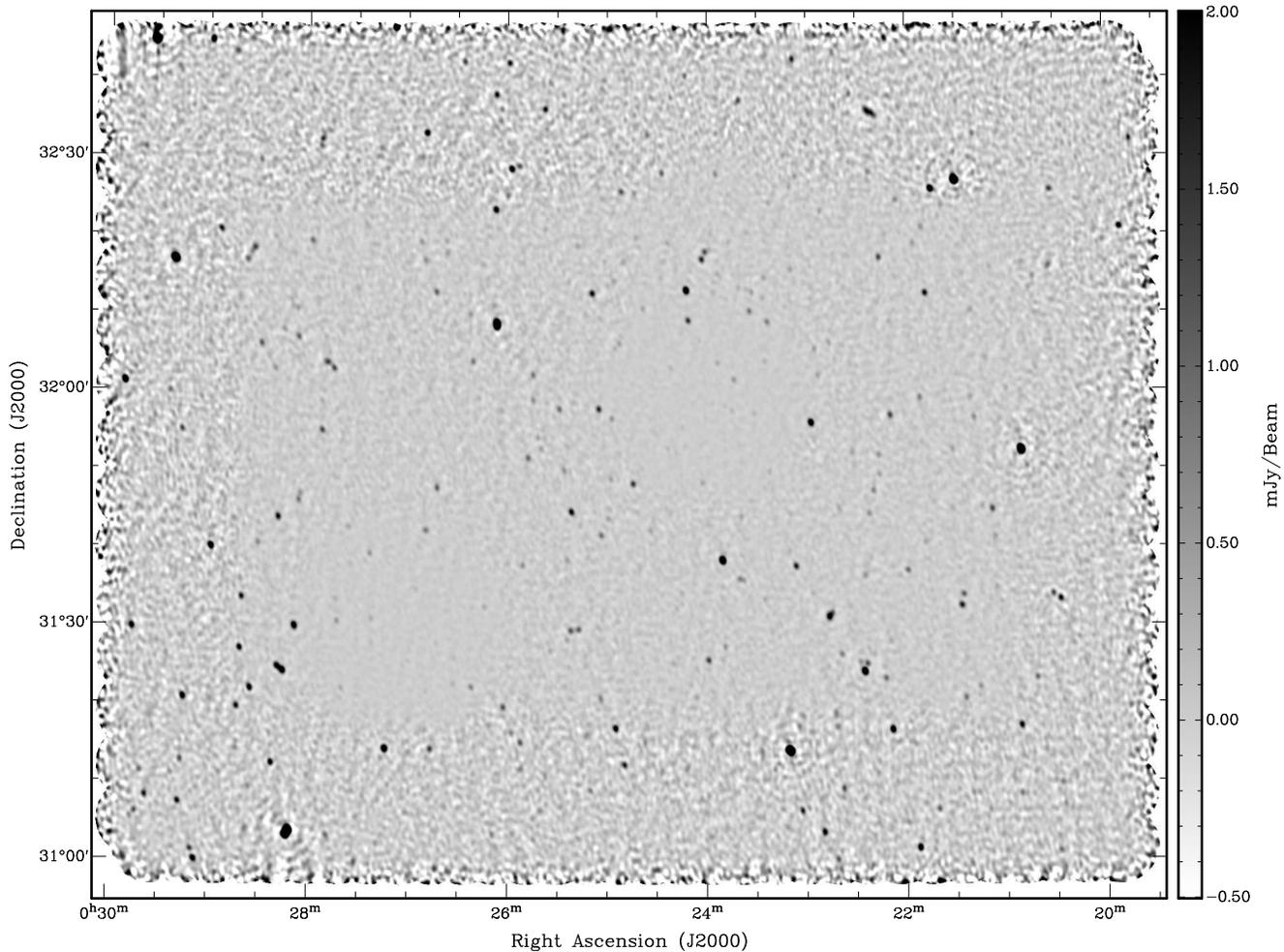}}
  \caption{The raster map of one of the 10C survey fields
           (J0024$+$3152).  The map was produced by combining
           approximately 1000 individual constituent maps,
           weighted according to their RMS noises; the individual
           maps were \textsc{clean}ed before being combined.  The
           area of lower noise in the centre of the raster map is
           clearly visible.}
  \label{fig:example_map}
\end{figure*}

The techniques used for observing, mapping and source extraction
are described fully in Paper~I.  The fields, the positions of which
are given in Section~\ref{Survey Fields}, were surveyed using a
`rastering' technique, with observations being carried out between
2008 August and 2010 June.  Each field was observed using a set of
telescope pointing directions spaced at 4.0~arcmin intervals and
lying on a 2-D hexagonally-gridded lattice, projected on to the
plane of the sky.  A raster map that combines the individual
\textsc{clean}ed maps belonging to each of the pointing directions
was produced for each field; the raster map for one of the survey
fields is shown in Fig.~\ref{fig:example_map}.  In addition, a
noise map that shows how the noise varies across the raster map was
created for each field; these noise maps are used in identifying
sources, as described in Paper~I.

Information about the sources was extracted from the raster maps,
using a combination of in-house software and tasks belonging to the
\textsc{aips}\footnote{\textsc{astronomical image processing system} -- www.aips.nrao.edu/},
for inclusion in the 10C source catalogue.  Source-finding was
carried out using a flux-density threshold of $4.62 \sigma$; the
reason for this slightly unusual choice is explained in
Sections~\ref{Checking the flux-density scale} and
\ref{Correcting the sources' flux densities for phase errors}.

In Section~\ref{Survey Fields}
areas within each of the survey fields,  bounded by lines of
constant right ascension and declination, complete to flux
densities of 1.0 and 0.5~mJy, are defined.
A short section of the catalogue is shown in Table~\ref{tab:landfig}.
The methods used to extract the various parameters are described
fully in Paper~1; the final column indicates which of the areas
(see Section~\ref{Survey Fields}) each of the sources lies within.
The complete source list, which contains 1897 entries, is available
online at www.mrao.cam.ac.uk/surveys/10C.  It is eventually planned
to also make the \textsc{fits} maps for each of the survey fields
publicly available on this website.

Individual positional error estimates have not been assigned to
each source.  The positional errors in both RA and Dec. for a
source detected at the 5-$\sigma$ level, in any of the survey
fields, are estimated to be 3--4~arcsec; this range reflects the
fact that the synthesised beam is slightly elliptical and has
dimensions which vary with field declination.  The assessment of
the source positional accuracy was made using simulations in which
point sources were inserted into the map of one of the survey
fields; the extracted source positions were compared with the
nominal values.

The results of the simulation were found to agree well with the
positional errors that would be expected from theory, taking
account of Gaussian thermal noise.  Higher-resolution follow-up
observations are required to assess the positional accuracy for the
brighter sources.  Only the highest-flux-density source in the 10C
survey catalogue has a counterpart in the catalogue of the
Jodrell-VLA Astrometic Survey
\citep[JVAS;][]{patnaik1992,browne1998,wilkinson1998}.  The
measured positions of this source agree within 3.5~arcsec.

There were some problems with the analysis and mapping of the 10C
survey data that required some adaptations to the source-finding
procedure described in Paper~I.  These adaptations are described
in the remainder of this section.

\begin{table*}
\vbox to220mm{\vfil Landscape table to go here.
\caption{}
\label{tab:landfig}
\vfil}
\end{table*}

\subsection{Checking the flux-density scale}\label{Checking the flux-density scale}

As a check of the raster maps' flux-density scale, $\approx 50$ of
the brightest sources detected in the maps were selected for
pointed follow-up observations, carried out during 2010 June
and August using the LA.  To avoid complicating the analysis, only
sources that showed no evidence of extension in the raster maps
were selected.  The data from these observations were mapped using
the same \textsc{clean}ing scheme used for the raster maps but
phase self-calibration was also applied.

\begin{figure}
 \begin{center}
  \includegraphics[scale=0.4,angle=270]{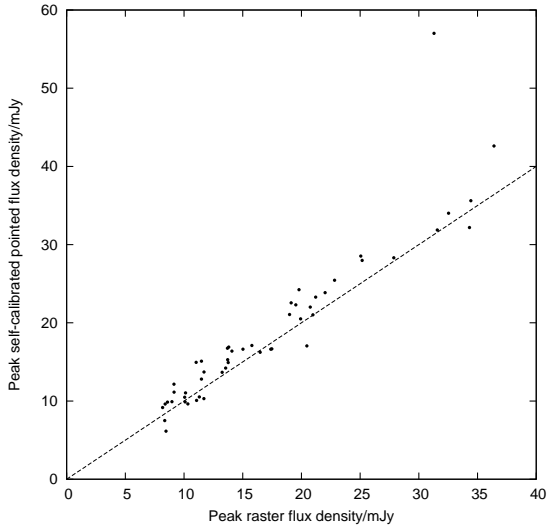}
  \caption{The peak flux densities of $\approx 50$ sources as
           measured from the raster maps and from maps created
           using self-calibrated data from pointed follow-up
           observations.  Points lying on the line represent
           sources having identical flux densities when measured
           using the raster and pointed maps.}
  \label{fig:pointed_vs_raster}
 \end{center}
\end{figure}

Fig.~\ref{fig:pointed_vs_raster} shows the peak flux densities for
each of the sources measured using the raster and
(self-calibrated) pointed maps.  The figure indicates that the flux
densities of the sources measured from the raster maps are
systematically low compared with those from the self-calibrated
pointed observations.  However, because the pointed follow-up
observations were carried out about 2~yr after the commencement
of the raster observations, it is important to consider the effect
of source variability on this result.

Fig.~\ref{fig:pointed_vs_raster} indicates that the pointed flux
density of one source is almost twice its raster flux density;
this difference is almost certainly attributable to genuine
flux-density variability.  Having said this, the number of
genuinely variable sources within the sample is likely to be
small.  The sources selected for pointed follow up have flux
densities ranging between approximately 10 and 40 mJy.
Results from \citet{waldram2010} indicate that 15-GHz-selected
samples containing sources with flux densities in this range are
likely to be dominated by steep-spectrum sources, which do not
ordinarily display significant variability.

Nevertheless, the median percentage difference has been used to
quantify the discrepancy between the pointed and raster flux
densities because it is less sensitive to genuine source
variability than the mean, the value of which could be strongly
affected by a small number of highly-variable sources within the
sample.  The median percentage difference between the pointed and
raster flux densities was calculated to be 8.2~per~cent with an
uncertainty in this value of  $\approx 2$~per~cent.

Since a number of the 10C survey fields overlap with areas mapped
as part of the 9C survey, as an additional check the flux densities
measured from the pointed observations were compared with the
values measured from pointed 9C observations.  Having corrected the
phase errors in the 10C pointed observations using
self-calibration, there was found to be good agreement between the
9C and 10C pointed values with a median percentage difference of
$< 1$~per~cent -- the median is again used for the same reason as
given above.  It is noted that, owing to the small difference in
the observing frequencies and the typical spectra of radio sources,
the 10C flux densities might be expected to be slightly lower
than the 9C values -- this was, in fact, the case.

Data from a large number of observations, carried out during a
range of weather conditions, were combined to create the raster
maps; whereas the data used to produce the pointed maps were
collected during single, short observations during relatively good
(dry) observing conditions.  As a result, the weather conditions
during the pointed observations may not be typical of those during
the raster observations.  However, in practice data collected
during periods of poor weather are either omitted entirely or
significantly downweighted with respect to data collected during
good weather conditions.

A modulated noise signal, injected at the front end of each
antenna, was monitored throughout the raster observations.  The
data were then weighted, on a sample-by-sample basis, according to
the value of the `rain gauge' -- that is the ratio of the power of
the modulated noise signal to the total power input to the
correlator (which is kept constant) to that obtained in cool, dry,
clear weather conditions.  In addition, data for which the rain
gauge was less than 50~per~cent (this is rather a conservative
criterion) were omitted.  This is because the amplitude correction
applied to the data becomes unreliable during heavy rain, as
no attempt to account for atmospheric absorption is made in
applying this correction.

Atmospheric-related phase effects are not thought to make any
significant contribution to the phase errors present in the data.
During LA observations a phase-calibrator source is observed for
1~min at a time at 10~min intervals.  Even using the telescope's
longest (110-m) baseline, the measured phase varies only slowly on
timescales much longer than 10~min.  If the measured phase for a
calibrator source does change by more than $30^{\circ}$ between
successive visits to the source, the affected data are
automatically omitted during the data-reduction stage; however,
such a large phase difference between successive calibrator visits
is observed only very rarely.  Similarly, any data for which the
estimated error in the measured phase of the calibrator is
$> 15^{\circ}$ are also omitted automatically.

The discrepancy between the flux densities measured using the
phase-self-calibrated pointed observations and the raster
observations is instead attributed to phase errors in AMI data
resulting from the uneven spacings of the time delays in the
telescope's lag correlator. \citet{holler2007}  explain this problem
in detail and propose a solution.  In practice, however, further
work is required to analyse the data from the correlator correctly.

\subsection{Correcting the sources' flux densities for phase errors}\label{Correcting the sources' flux densities for phase errors}

Unfortunately, the great majority of the sources in the 10C raster
maps are detected with insufficient signal-to-noise to allow
self-calibration to be successfully applied to the data.
Therefore, a correction to the source flux densities based on the
difference between the flux densities measured from the raster
maps and the self-calibrated pointed maps is applied; as a final
step in the source-finding procedure, all source flux densities are
multiplied by 1.082 before inclusion in the 10C catalogue.

The uncertainties in the flux densities are increased to take
account of this scaling and the uncertainty in the correction
factor ($\approx 2$~per~cent), which is in any case small compared
with the estimated LA calibration uncertainty of 5 per cent.  As in
Paper~I, the uncertainty in an uncorrected peak flux density, $S$,
is taken to be $\sqrt{\sigma_{\mathrm{n}}^{2} + (0.05S)^{2}}$,
where $\sigma_{\mathrm{n}}$ is the thermal noise at the source
position, estimated from the noise map.  The uncertainty in the
corrected value is, therefore,
\begin{eqnarray}
1.082S \sqrt{ \frac{\sigma_{\mathrm{n}}^{2} + (0.05S)^{2}}{S^{2}} + 0.02^{2}}.
\end{eqnarray}

Initially, it was intended to carry out source finding at
$5 \sigma$; since, assuming Gaussian statistics for the map noise,
such a scheme would result in a highly reliable catalogue with
$\approx 0.1$ false detections.  Further, the completeness of such
a catalogue would have been very high ($\gtrsim 94$ per cent) at
0.5~mJy and 1~mJy in the deep and shallow regions respectively.
However, owing to the effect of phase errors, a source of
$S = 1$~mJy falling within the shallow area of the survey, which
ought to be detected at $\geq 5 \sigma$, will only be detected at 
$\geq 5\sigma/1.082 = 4.62 \sigma$.

Therefore, in order to achieve the desired high levels of
completeness at 0.5~mJy and 1~mJy, it was decided to carry out the
source finding at $4.62 \sigma$.  The catalogue completeness is
discussed fully in Section~\ref{Completeness}.  The slight
relaxation in the source-finding criterion is likely to have only a
small adverse effect on the reliability of the catalogue.  Assuming
Gaussian statistics for the map noise, and given the number of
synthesised beam areas in the survey maps, it is estimated that
about one source in the final catalogue will be a false positive.

\subsection{Excluding areas around bright sources}\label{Excluding areas around bright sources}

The 10C raster maps often display an increased level of noise
around bright ($\gtrsim 15~\mathrm{mJy}$) sources.  This is
attributable to amplitude, phase and deconvolution errors in the
data.  The elevated noise level close to bright sources is
generally not fully reflected by the noise maps, because the noise
is highly non-Gaussian in these regions.  Therefore, such
detections close to sources of $S > 15$~mJy were automatically
excluded from the final source catalogue.  In addition, detections
close to a number of fainter sources (the faintest being
$\approx 9$~mJy) were also excluded manually from the final
catalogue after inspecting the maps.

Empirically, the area of elevated noise around each bright source
(the `exclusion zone'), from which the source detections were
rejected, was found to be well-represented by a circle, centred on
the source position, of radius
\begin{eqnarray}
\label{eqn:ex_zone}
r_{\mathrm{e}} = 12\left(\frac{S_{\mathrm{pk}}/\mathrm{mJy}}{250}\right)^{1/2}\mathrm{arcmin}
\mathrm{,}
\end{eqnarray}
where $S_{\mathrm{pk}}$ is the peak flux density of the source.
Table~\ref{tab:ex_zones} shows the centre coordinates and radii of
the exclusion zones applied to the survey data.

\begin{table}
 \begin{center}
 \caption{The centre positions of the exclusion zones around
          bright sources and their radii.}
 \label{tab:ex_zones}
 \begin{tabular}{@{}c c d }
 \hline
 RA          & Dec.            & \dhead{Radius (arcmin)} \\
 \hline
 00:20:50.4  & $+$31:52:29     &   3.39                  \\
 00:21:29.8  & $+$32:26:60     &   3.63                  \\
 00:23:09.9  & $+$31:14:01     &   4.26                  \\
 00:26:06.2  & $+$32:08:33     &   2.83                  \\
 00:28:10.7  & $+$31:03:46     &   3.85                  \\
 00:29:20.4  & $+$32:16:55     &   3.17                  \\
 00:29:33.1  & $+$32:44:58     &   4.45                  \\
 02:59:55.1  & $+$26:27:26     &   2.31                  \\
 03:01:05.5  & $+$25:47:16     &   2.92                  \\
 03:01:37.3  & $+$25:41:54     &   3.04                  \\
 07:31:17.4  & $+$53:38:58     &   4.25                  \\
 07:36:52.9  & $+$54:29:17     &   2.70                  \\
 08:18:16.1  & $+$69:16:53     &   4.06                  \\
 08:23:02.5  & $+$69:14:20     &   2.94                  \\
 09:35:59.5  & $+$31:27:27     &   3.17                  \\
 09:36:36.9  & $+$32:03:35     &   3.31                  \\
 09:37:06.2  & $+$32:06:58     &   5.41                  \\
 09:41:03.2  & $+$31:26:14     &   3.30                  \\
 09:41:46.2  & $+$31:55:03     &   3.05                  \\
 09:42:08.8  & $+$32:06:42     &   3.48                  \\
 10:47:19.3  & $+$58:21:14     &   4.93                  \\
 10:49:40.0  & $+$58:35:31     &   3.35                  \\
 10:50:07.1  & $+$56:53:37     &   3.38                  \\
 10:50:54.0  & $+$58:32:33     &   3.50                  \\
 10:51:41.4  & $+$59:13:08     &   3.80                  \\
 10:52:25.4  & $+$57:55:08     &   3.46                  \\
 10:52:54.5  & $+$59:22:18     &   3.56                  \\
 10:54:26.9  & $+$57:36:48     &   3.69                  \\
 15:20:41.6  & $+$44:13:18     &   3.43                  \\
 15:21:49.4  & $+$43:36:37     &  12.51                  \\
 15:27:51.8  & $+$43:52:05     &   2.85                  \\
 15:28:19.8  & $+$42:33:35     &   4.01                  \\
 15:40:33.5  & $+$44:34:01     &   3.32                  \\
 15:41:10.0  & $+$44:56:34     &   4.58                  \\
 15:42:23.1  & $+$43:59:15     &   3.81                  \\
 15:46:04.5  & $+$44:49:14     &   3.02                  \\
 17:25:34.5  & $+$41:53:03     &   4.33                  \\
 17:27:49.3  & $+$42:21:40     &   4.45                  \\
 17:29:01.9  & $+$41:40:04     &   2.68                  \\
 17:30:41.6  & $+$41:02:58     &   6.14                  \\
 17:31:23.7  & $+$41:01:38     &   3.01                  \\
 17:37:59.6  & $+$41:54:51     &   5.00                  \\
 17:38:35.4  & $+$42:21:43     &   3.08                  \\
 17:40:08.9  & $+$41:36:09     &   6.31                  \\
 17:40:17.2  & $+$42:14:30     &   2.93                  \\
 17:40:52.1  & $+$42:34:47     &   5.79                  \\
 \hline
 \end{tabular}
 \end{center}
\end{table}

The dynamic range of the LA -- that is the ratio of a source's peak
flux density to the flux density of the brightest artefact in a map
close to the source -- is $\approx$~50:1.  The brightest source
detected in the 10C survey is $\approx 270$~mJy. A conservative
approach, which also serves to simplify the completeness areas for
construction of the source count, has been taken by assuming that
sources with $S \geq 9$~mJy can be detected anywhere in the total
areas.  In contrast, any putative source with a peak flux density
less than this value that falls within the exclusion zone of a
bright source is not included in the final 10C source catalogue.
Thus, the area used for calculating the source counts for sources
with $S < 9$~mJy does not include the exclusion zones around bright
sources.  The total area excluded from around bright sources is
0.6~deg$^{2}$.

\section{Survey Fields}\label{Survey Fields}

\begin{figure}
 \begin{center}
  \includegraphics[scale=0.40,angle=270]{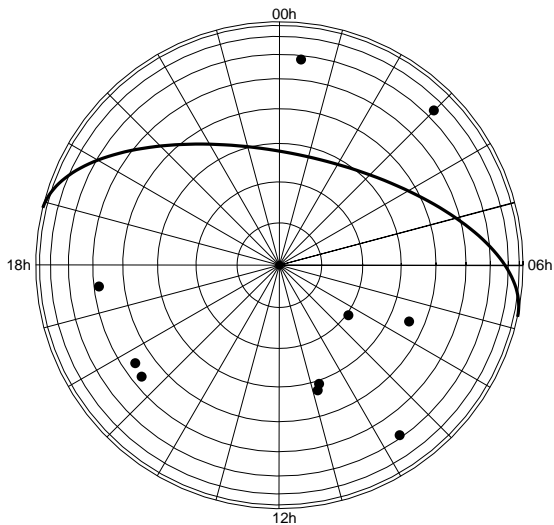}
  \caption{The positions of the 10C survey fields, shown using an
           equatorial-plane projection with the North Pole at the
           centre.  The declination circles are at intervals of
           10$^{\circ}$ and the Galactic plane is indicated by the
           thick black line.}
  \label{fig:10C_fields}
 \end{center}
\end{figure}

The 10C survey fields are centred at J0024+3152, J0259+2610,
J0734+5432, J0824+6931, J0939+3115, J1046+5904, J1052+5730,
J1524+4321, J1543+4420 and J1733+4148.  Fig.~\ref{fig:10C_fields}
shows the positions of these fields, which were chosen to be widely
spread in RA and away from the Galactic plane.  For each field, areas
complete to 1.0 and 0.5~mJy (apart from the exclusion zones) have
been defined by selecting those areas in which the noise,
$\sigma_{\mathrm{n}}$, estimated using the relevant noise map
is $0.1 \leq \sigma_{\mathrm{n}} < 0.2$~mJy and
$\sigma_{\mathrm{n}} < 0.1$~mJy respectively.  Over a large portion
of the areas, the estimated noise is significantly lower than the
upper bounds and, therefore, the completeness is close to 100 per
cent for both regions at their nominal completeness levels.

\begin{figure}
 \epsfig{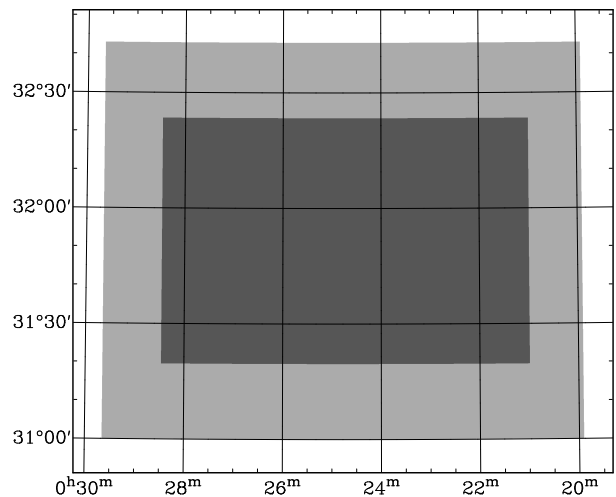}\qquad\qquad
 \caption{The shallow (lighter shading) and deep areas belonging to
          J0024+5432.}
  \label{fig:10C_areas}
\end{figure}

The areas complete to 0.5~mJy are entirely contained within the areas
complete to 1.0~mJy and are referred to as the `deep' regions.  Areas
of higher noise, complete to 1.0~mJy, but excluding the deep regions,
are referred to as the `shallow' areas.  Fig.~\ref{fig:10C_areas}
shows the deep and shallow regions for one of the fields.  The
survey catalogue includes a flag for each source indicating whether
it falls in the deep (D) or shallow (S) areas.  Sources that fall
outside these regions altogether (i.e. in areas of higher noise
towards the edges of the raster maps) are indicated by `N'.  For
a source with evidence of extension the flag is based on the source's
centroid position, otherwise the peak position is used.  The `total'
areas are those areas complete to 1~mJy but not excluding the deep
regions -- in other words, the combined deep and shallow areas.
The lines of right ascension and declination bounding the total and
deep areas, for each of the fields, are given in Tables
\ref{tab:total_areas} and \ref{tab:deep_areas} respectively.

\begin{table*}
 \begin{minipage}{10cm}
 \begin{center}
  \caption{The areas complete to 1~mJy.}
  \label{tab:total_areas}
  \begin{tabular}{@{}c c c c d }
   \hline
     Field      & RA range                 & Dec. range                 & \dhead{Area (deg$^{2}$)} \\
   \hline
     J0024+3152 & 00:19:54.2 to 00:29:38.3 & $+$31:00:04 to $+$32:43:05 & 3.56                     \\
     J0259+2610 & 02:56:43.6 to 03:02:32.8 & $+$25:19:17 to $+$27:02:17 & 2.24                     \\
     J0734+5432 & 07:30:02.8 to 07:38:50.7 & $+$53:41:42 to $+$55:23:07 & 2.16                     \\
     J0824+6931 & 08:17:08.6 to 08:31:23.8 & $+$68:41:04 to $+$70:22:41 & 2.11                     \\
     J0939+3115 & 09:36:22.1 to 09:42:25.9 & $+$30:24:22 to $+$32:06:18 & 2.20                     \\
     J1046+5904 & 10:39:25.5 to 10:52:31.2 & $+$58:13:12 to $+$59:55:07 & 2.86                     \\
     J1052+5730 & 10:47:36.8 to 10:57:21.3 & $+$56:48:56 to $+$58:10:22 & 1.78                     \\
     J1524+4321 & 15:20:10.7 to 15:29:24.6 & $+$42:30:20 to $+$44:11:21 & 2.83                     \\
     J1543+4420 & 15:38:30.5 to 15:47:46.6 & $+$43:28:43 to $+$45:11:13 & 2.83                     \\
     J1733+4148 & 17:25:37.0 to 17:41:01.3 & $+$40:57:34 to $+$42:40:06 & 4.90                     \\
   \hline
  \end{tabular}
 \end{center}
 \end{minipage}
\end{table*}

\begin{table*}
 \begin{minipage}{10cm}
 \begin{center}
  \caption{The areas complete to 0.5~mJy.  Note that there are three deep regions
           associated with J0259+2610.}
  \label{tab:deep_areas}
  \begin{tabular}{@{}c c c d }
   \hline
     Field      & RA range                 & Dec. range                 & \dhead{Area (deg$^{2}$)} \\
   \hline
     J0024+3152 & 00:20:59.1 to 00:28:27.1 & $+$31:19:38 to $+$32:23:29 & 1.69                     \\
     J0259+2610 & 02:57:50.5 to 03:01:24.1 & $+$25:38:31 to $+$26:43:51 & 0.87                     \\
                & 02:56:50.2 to 02:57:50.5 & $+$25:53:05 to $+$26:28:22 & 0.13                     \\
                & 03:01:24.1 to 03:02:31.4 & $+$25:53:05 to $+$26:28:22 & 0.15                     \\
     J0734+5432 & 07:31:47.0 to 07:37:06.6 & $+$54:00:59 to $+$55:04:59 & 0.82                     \\
     J0824+6931 & 08:19:21.5 to 08:29:26.1 & $+$68:57:47 to $+$70:07:19 & 1.02                     \\
     J0939+3115 & 09:37:32.8 to 09:41:15.9 & $+$30:42:52 to $+$31:47:13 & 0.85                     \\
     J1046+5904 & 10:41:30.0 to 10:50:24.5 & $+$58:32:47 to $+$59:35:08 & 1.19                     \\
     J1052+5730 & 10:49:57.7 to 10:55:32.7 & $+$57:08:03 to $+$57:51:28 & 0.54                     \\
     J1524+4321 & 15:21:32.3 to 15:27:57.3 & $+$42:50:56 to $+$43:52:21 & 1.19                     \\
     J1543+4420 & 15:39:51.6 to 15:46:21.3 & $+$43:49:49 to $+$44:50:48 & 1.18                     \\
     J1733+4148 & 17:26:44.0 to 17:39:49.8 & $+$41:17:01 to $+$42:19:40 & 2.54                     \\
   \hline
  \end{tabular}
 \end{center}
 \end{minipage}
\end{table*}

\section{Completeness}\label{Completeness}

\subsection{Simulations}\label{Simulations}

Simulations were carried out to investigate the completeness of the
survey.  A number of realisations were used (twelve for the deep
areas and thirteen for the shallow) in which 250
equal-flux-density, simulated point sources were inserted into the
raster map of J0024+3152.  The flux density of the simulated
sources was different for each of the realisations.  The positions
of the simulated sources were chosen randomly but were not altered
between realisations; to avoid the simulated sources affecting
each other, it was insisted that no simulated source could lie
within 2~arcmin of any other.  Sources were not inserted in the
exclusion zones around bright sources, since, as explained above,
the completeness limit is expected to be much higher for these
areas compared with the remainder of the maps.  In order to
investigate the completeness as a function of flux density, the
flux density of the sources was changed between each realisation.

For each realisation, the ordinary source-finding procedures were
applied to the map and the proportions of simulated sources that
were recovered was calculated for the shallow and deep areas.
Three of the simulated sources were found to lie too close to real
sources to be detected separately.  In these cases, the simulated
source was considered to be detected if the recovered source
position was closer to the simulated rather than the real source
position.

\begin{figure}
 \begin{center}
  \includegraphics[scale=0.9,angle=0]{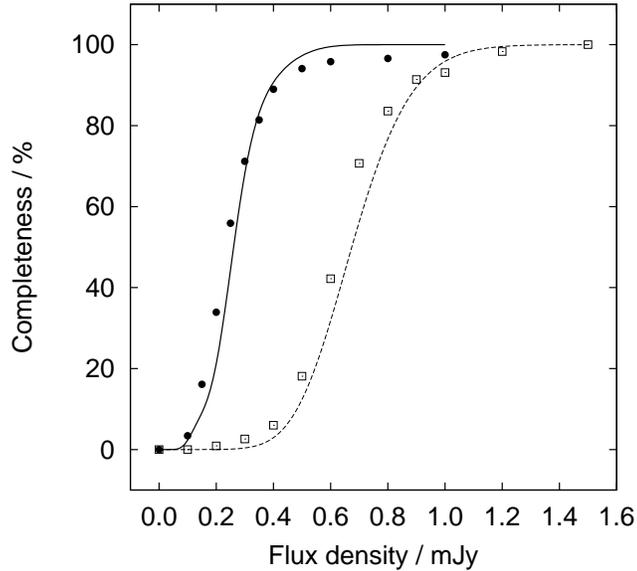}
  \caption{Results of a simulation to investigate the
           completeness of the survey.  Filled circles show the
           proportion of the simulated sources recovered as a
           function of flux density within the deep
           ($\sigma_{\rm{n}} \leq 0.1$~mJy) area of J0024+3152.
           The solid line shows the completeness predicted based on
           the noise-map pixel values assuming Gaussian statistics
           for the noise.  The open squares and dashed line show
           the results for the shallow
           ($0.1 < \sigma_{\rm{n}} \leq 0.2$~mJy) area.}
  \label{fig:completeness_sim}
 \end{center}
\end{figure}

Fig.~\ref{fig:completeness_sim} shows the proportion of sources
detected for each of the realisations compared with the expected
detection rate, which is calculated for a particular region of the
map as follows.  For a specific source flux density, the probability
of detection can be calculated at the position of each pixel by
taking account of the value of the noise map at that position and
by using Gaussian statistics.  The probability of detection over the
whole region is straightforwardly calculated by averaging over pixels.

The plots show that, in general, there is good
agreement between the predicted completeness curves and the results
of the simulation.  However, for the fainter sources, the detection
rate is slightly higher than predicted, whilst the converse is true
for the brighter sources.  Source confusion is likely to be
responsible for these results.  A number of the highest
flux-density simulated sources, that would otherwise have been
detected, were not recovered owing to their proximity to real,
bright sources.  This effect prevented the completeness from
reaching 100 per cent as quickly as predicted.  At fainter flux
densities, however, the opposite effect, whereby a source is
boosted in flux density owing to its proximity to a real, fainter
source, so that it is unexpectedly detected, becomes important.

The peak flux densities of sources detected with low
signal-to-noise ratios (SNRs) are typically biased slightly high,
because the peak position tends to be coincident with a positive
noise fluctuation.  This is an additional factor serving to boost
the detection rate at low SNRs.  The presence of this effect was
confirmed by extracting the flux densities of the sources at the
precise positions with which they were simulated; the values were
found to be systematically low compared with the extracted
\textit{peak} flux densities and were found to reflect better the
simulated flux densities.

\subsection{Using the noise maps to estimate the completeness}
\label{Using the noise maps to estimate the completeness}

Having established by simulation that, assuming Gaussian statistics
for the noise, the noise maps can be used to provide reasonable
estimates of the survey completeness, the noise maps from all the
fields were used to estimate the completeness of the 10C survey.
The probability of a source of true flux density $\hat{S}$ being
included in the survey catalogue when located on a pixel with a
corresponding noise-map value of $\sigma_{\rm{n}}$ was taken,
according to Gaussian statistics, to be
\begin{eqnarray}
\label{eqn:completeness}
P(\hat{S} \geq 4.62 \sigma_{\rm{n}}) = \int^{\infty}_{4.62\sigma_{\rm{n}}} \frac{1}{\sqrt{2\pi\sigma_{\rm{n}}^{2}}}\exp{-\frac{\left(x - \hat{S}/1.082\right)^{2}}{2\sigma_{\rm{n}}^{2}}}~\mathrm{d}x.
\end{eqnarray}
In carrying out this calculation the fact that sources are detected
with flux densities lower than their true values has been taken
into account by including the factor of $1/1.082$.

\begin{figure}
 \begin{center}
  \includegraphics[scale=0.40,angle=270]{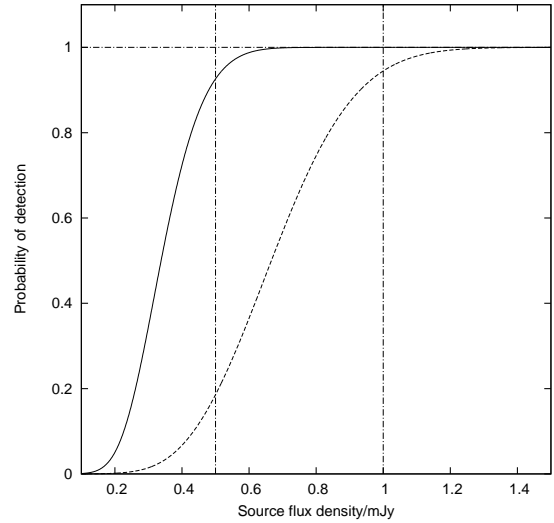}
  \caption{The estimated probability of detection for the shallow
           (dashed line) and deep (solid line) areas for all survey
           fields.  The dot-dashed horizontal line indicates a probability of
           one.  The dot-dashed vertical lines indicate the 0.5-mJy
           and 1.0-mJy nominal completeness limits for the deep and
           shallow areas respectively.}
  \label{fig:completeness_curve}
 \end{center}
\end{figure}

Knowing the actual distribution of noise-map pixel values,
Equation~\ref{eqn:completeness} can be used to estimate the
completeness, as a function of flux density, for the shallow and
deep regions of the survey; Fig.~\ref{fig:completeness_curve}
shows these estimates for both areas.  For the shallow regions, the
catalogue is estimated to be $\approx 94$-per-cent complete by
1~mJy and 99-per-cent complete by $\approx 1.16~$mJy.  For the
deep areas, the catalogue is estimated to be
$\approx 93$-per-cent complete by 0.5~mJy and 99-per-cent complete
by $\approx 0.61$~mJy.

\section{Source counts}\label{Source counts}

The 15.7-GHz differential source counts have been calculated by
binning the sources from the final catalogue according to their
peak flux densities, except for those that display evidence of
being extended relative to the telescope synthesised beam (this is
the case for 5.5~per~cent of sources), for which integrated flux
densities were used.

The binned differential-source-count data are given in
Table~\ref{tab:10C_count}.  For the highest flux-density bin, data
from the entirety of the total areas have been used.  The bins for
sources with flux densities between 1 and 9~mJy include sources
from the total regions but exclude the areas around bright sources
given in Table~\ref{tab:ex_zones}.  The bins for sources with flux
densities between 0.5 and 1~mJy include sources from the deep
regions (again excluding areas around bright sources).    The
source count is not calculated for $S > 25$~mJy, since it is biased
low in this flux-density range; using 9C data, several of the
fields were selected to contain as few sources with $S > 25$~mJy
as possible.

\begin{table}
 \begin{center}
  \caption{Data for the 10C source count.}
  \label{tab:10C_count}
  \begin{tabular}{@{}d d d d }
   \hline
     \dhead{Bin start}  & \dhead{Bin end}   &  \dhead{Number of} & \dhead{Area}        \\
     \dhead{$S$ (mJy)}  & \dhead{$S$ (mJy)} &  \dhead{sources}   & \dhead{(deg$^{2}$)} \\
   \hline
   9.000                & 25.000            &  46                & 27.46               \\
   5.500                & 9.000             &  51                & 26.86               \\
   2.900                & 5.500             &  142               & 26.86               \\
   2.050                & 2.900             &  135               & 26.86               \\
   1.500                & 2.050             &  148               & 26.86               \\
   1.250                & 1.500             &  113               & 26.86               \\
   1.000                & 1.250             &  160               & 26.86               \\
   0.900                & 1.000             &  36                & 11.96               \\
   0.775                & 0.900             &  56                & 11.96               \\
   0.680                & 0.775             &  51                & 11.96               \\
   0.600                & 0.680             &  64                & 11.96               \\
   0.540                & 0.600             &  61                & 11.96               \\
   0.500                & 0.540             &  46                & 11.96               \\
   \hline
  \end{tabular}
 \end{center}
\end{table}

At 0.5~mJy, the completeness limit of the deep areas, the survey
is limited by thermal rather than confusion noise.  Above this
flux-density level, it is estimated that there are typically
170 LA synthesised beam areas per source.

The effect of the calibration errors of $\approx 5$~per~cent is
negligible serving to boost the number of sources in each bin by
$<< 1$~per~cent (owing to the sign of the slope of the source
count).  However, bias \citep{eddington1913} due to the thermal
noise will play a more important role for the bins at the faintest
flux-density levels.  Given the slope of the counts and the noise
properties of the deep region, the number of sources in the
faintest flux-density bin is expected to be boosted by
$\approx 7$~per~cent.  However, this effect is almost exactly
balanced by the small degree of incompleteness that affects these
faintest flux-density bins.  Consequently, no corrections for the
effect of Eddington bias or incompleteness have been applied.

\begin{figure}
 \begin{center}
  \includegraphics[scale=0.44,angle=270]{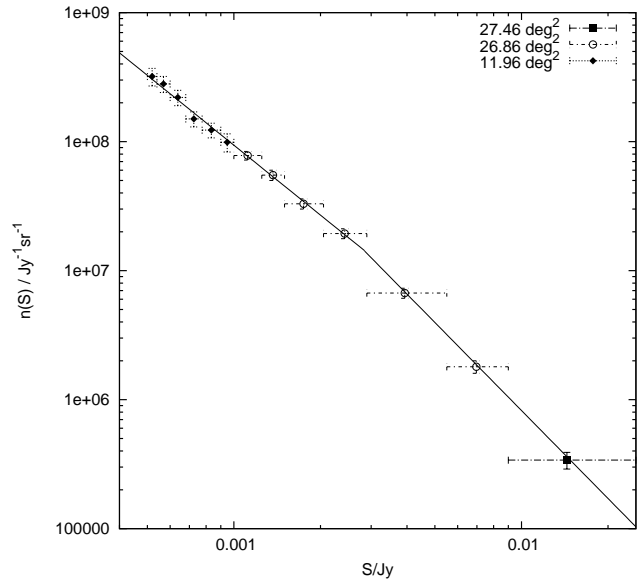}
  \caption{The 10C differential source count.  Poisson errors, on
           the number of sources in each bin, are indicated for
           each of the points.  The bars parallel to the
           flux-density axis represent the bin widths, not error
           bars.  The fitted broken-power-law count is indicated by
           the solid line.}
  \label{fig:10C_count}
 \end{center}
\end{figure}

The 10C differential source count is shown in
Fig.~\ref{fig:10C_count}.  As in all subsequent plots showing
source counts, Poisson errors, on the number of sources in each
bin, are indicated for each of the points and the bars parallel to
the flux-density axis represent the bin widths, not error bars.

An attempt was made to fit a single power law to the data.  However,
such a model did not appear to fit the data well.  Consequently, a
broken power law has been fitted to the data.  A method was used
whereby the sum of the squared differences between the measured and
predicted areas under the curve, over all bins, was minimised.  In
carrying out the minimisation, the points were weighted according to
their respective Poisson errors.  The positions of the points in the
$S$ direction within the bins have been plotted, in
Fig.~\ref{fig:10C_count}, on the basis of the fitted exponents to
reflect the `centre of gravity' of each bin.  The fitted
differential source count is
\[n(S) \equiv \frac{\mathrm{d}N}{\mathrm{d}S} \approx \left\{
\begin{array}{l l}
~~24 \left(\frac{S}{\mathrm{Jy}}\right)^{-2.27}~\mathrm{Jy^{-1}~sr^{-1}}~\mathrm{for}~2.8 \leq S \leq 25~\mathrm{mJy} \\
376 \left(\frac{S}{\mathrm{Jy}}\right)^{-1.80}~\mathrm{Jy^{-1}~sr^{-1}}~\mathrm{for}~0.5 \leq  S < 2.8~\mathrm{mJy.}
\end{array} \right. \]

Tests were carried out to check that the broken power-law model
does indeed provide an improved fit to the data, compared with
the single power-law model.  An F-test indicated that the null
hypothesis -- that the data are more likely to have been drawn
from the simpler model -- can be rejected at $> 99.9$~per~cent
confidence.  A comparison of the models using Akaike's
information criterion was similarly emphatic, indicating that
the broken power-law model is $\approx 1000$ times more likely
to be correct than the simpler model.

\begin{figure}
 \begin{center}
  \includegraphics[scale=0.44,angle=270]{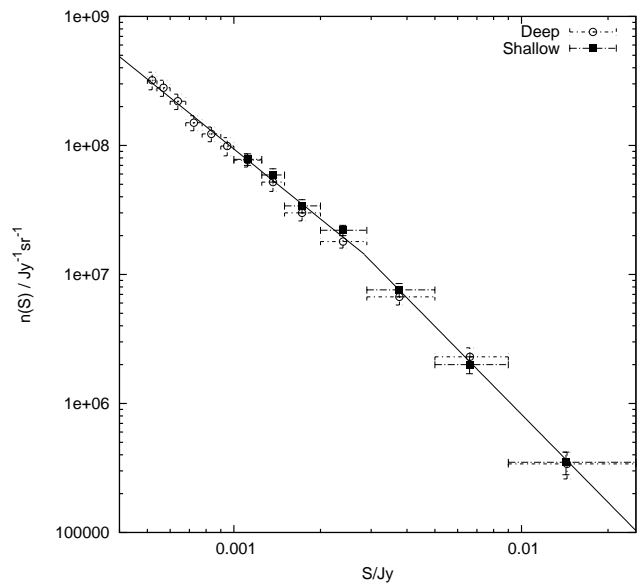}
  \caption{Differential source counts from the shallow and deep
           regions of the 10C survey.  The fitted 10C source
           count is indicated by the solid line.}
  \label{fig:10C_count_consistency_check}
 \end{center}
\end{figure}

In order to check the self-consistency of the source count, the
data were used to construct separate counts for the shallow and
deep regions.  Fig.~\ref{fig:10C_count_consistency_check} shows the
counts from the two regions overlaid.  The deep counts are derived
from an area amounting to 11.96~deg$^{2}$ (except for the highest
flux-density bin, which contains data from 12.19~deg$^{2}$).  The
shallow counts use data from 14.90~deg$^{2}$ (15.27~deg$^{2}$ for
the highest flux-density point).  The plot shows good agreement,
within the uncertainties, between the counts derived from the two
regions over the common flux-density range.

\subsection{Adding in data from the 9C survey}\label{Adding in data from the 9C survey}

It is possible to extend the 10C source count to higher flux
densities by the inclusion of data from the 9C survey.  It is also
possible to improve the source count statistics between 5.5 and
25~mJy by including 9C data, since the 9C survey contains regions
that are complete over this flux-density range.  There is a small
difference in the observing frequencies of the 9C (15.2~GHz) and
10C surveys (15.7~GHz).  Therefore, in combining the data sets, the
source flux densities from the 9C survey catalogue have been
corrected to take into account this difference.  The correction was
made by assuming a typical spectral index between 15.2 and 15.7~GHz
that varies as a function of source flux density.

\begin{figure}
 \begin{center}
  \includegraphics[scale=0.44,angle=270]{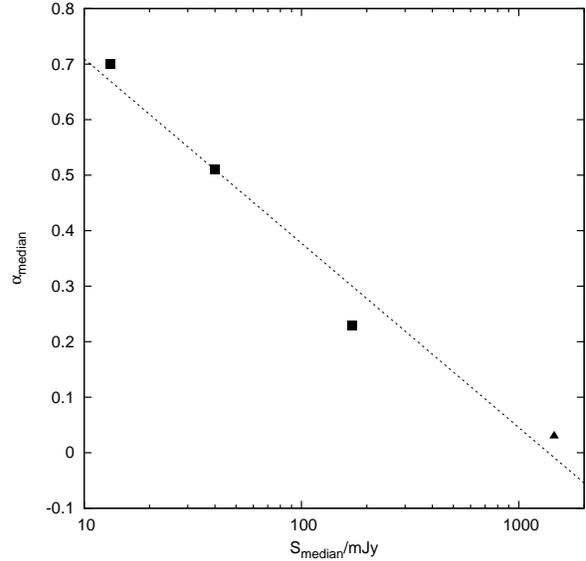}
  \caption{The filled squares represent the median 1.4-to-15.2-GHz
           spectral indices for complete 15.2-GHz-selected source
           samples from the 9C survey.  The filled triangle
           represents the median 16-to-33-GHz spectral index for
           a complete 33-GHz-selected sample from the \textit{WMAP}
           3-yr data.  The dashed line indicates the assumed
           typical spectral index, as a function of flux density,
           used to make corrections to the flux densities of
           individual 9C sources, to account for the small
           difference in the observing frequencies of the 9C and
           10C surveys.  This allowed the 10C source count
           to be extended to flux densities $> 25$~mJy by the
           inclusion of 9C data.  The typical spectral index was
           calculated by fitting a logarithmic function to the
           data points.}
  \label{fig:sindex_correction}
 \end{center}
\end{figure}

The assumed flux-density-dependent spectral index is indicated in
Fig.~\ref{fig:sindex_correction}.  The correction was calculated
by fitting a logarithmic function to the points in the plot.  For
each of the three lowest flux-density points (indicated by filled
squares), the median flux density versus the median value of the
1.4-to-15.2-GHz spectral index, $\alpha^{15.2}_{1.4}$, of sources,
with flux densities in the relevant ranges, detected as part of the
9C survey \citep[see Table~9 of][]{waldram2010} has been plotted.
For the highest flux-density point (indicated by the filled
triangle), the median flux density versus median value of
$\alpha_{16}^{33}$, measured by \citet{davies2009}, for sources
belonging to a flux-density-limited source sample
\citep{lopez-caniego2007} from the 3-yr
\textit{Wilkinson Microwave Anisotropy Probe} (\textit{WMAP}) data
has been plotted.  The number of sources belonging to the samples
represented by these points ranges between 84 and 381.

For both the 9C and \textit{WMAP} samples it has simply been
assumed that the  spectral indices can be extended over the small
additional frequency ranges to 15.7~GHz (from 15.2~GHz in the case
of the 9C samples and from 16~GHz for the \textit{WMAP} sample).
The approach followed is admittedly not perfect, particularly since
the sample for the highest flux-density point was selected at
33~GHz.  Nevertheless, because the difference in observing
frequencies between the surveys is very small, the corrections are
similarly small (at most a few percent) and the method is
considered acceptable.

\begin{table}
 \begin{center}
  \caption{Data for the combined 9C and 10C source counts.}
  \label{tab:joint_count}
  \begin{tabular}{@{}r r c r@{.} l }
   \hline
     \multicolumn{1}{c}{Bin start}  & \multicolumn{1}{c}{Bin end}   &  \multicolumn{1}{c}{Number of} & \multicolumn{2}{c}{Area}        \\
     \multicolumn{1}{c}{$S$ (mJy)}  & \multicolumn{1}{c}{$S$ (mJy)} &  \multicolumn{1}{c}{sources}   & \multicolumn{2}{c}{(deg$^{2}$)} \\
   \hline
   500.000   & 1000.000  &  8    & 520&00   \\
   200.000   & 500.000   &  27   & 520&00   \\
   100.000   & 200.000   &  47   & 520&00   \\
   60.000    & 100.000   &  92   & 520&00   \\
   40.000    & 60.000    &  97   & 520&00   \\
   30.000    & 40.000    &  99   & 520&00   \\
   25.000    & 30.000    &  79   & 520&00   \\
   16.000    & 25.000    &  62   & 124&60   \\
   12.000    & 16.000    &  64   & 124&60   \\
   10.000    & 12.000    &  48   & 124&60   \\
   9.000     & 10.000    &  15   &  47&83   \\
   6.400     & 9.000     &  48   &  47&23   \\
   5.500     & 6.400     &  44   &  47&23   \\
   2.900     & 5.500     &  142  &  26&86   \\
   2.050     & 2.900     &  135  &  26&86   \\
   1.500     & 2.050     &  148  &  26&86   \\
   1.200     & 1.500     &  140  &  26&86   \\
   1.000     & 1.200     &  133  &  26&86   \\
   0.900     & 1.000     &  36   &  11&96   \\
   0.775     & 0.900     &  56   &  11&96   \\
   0.680     & 0.775     &  51   &  11&96   \\
   0.600     & 0.680     &  64   &  11&96   \\
   0.540     & 0.600     &  61   &  11&96   \\
   0.500     & 0.540     &  46   &  11&96   \\
  \hline
  \end{tabular}
 \end{center}
\end{table}

\begin{figure*}
 \begin{center}
  \includegraphics[scale=0.88,angle=270]{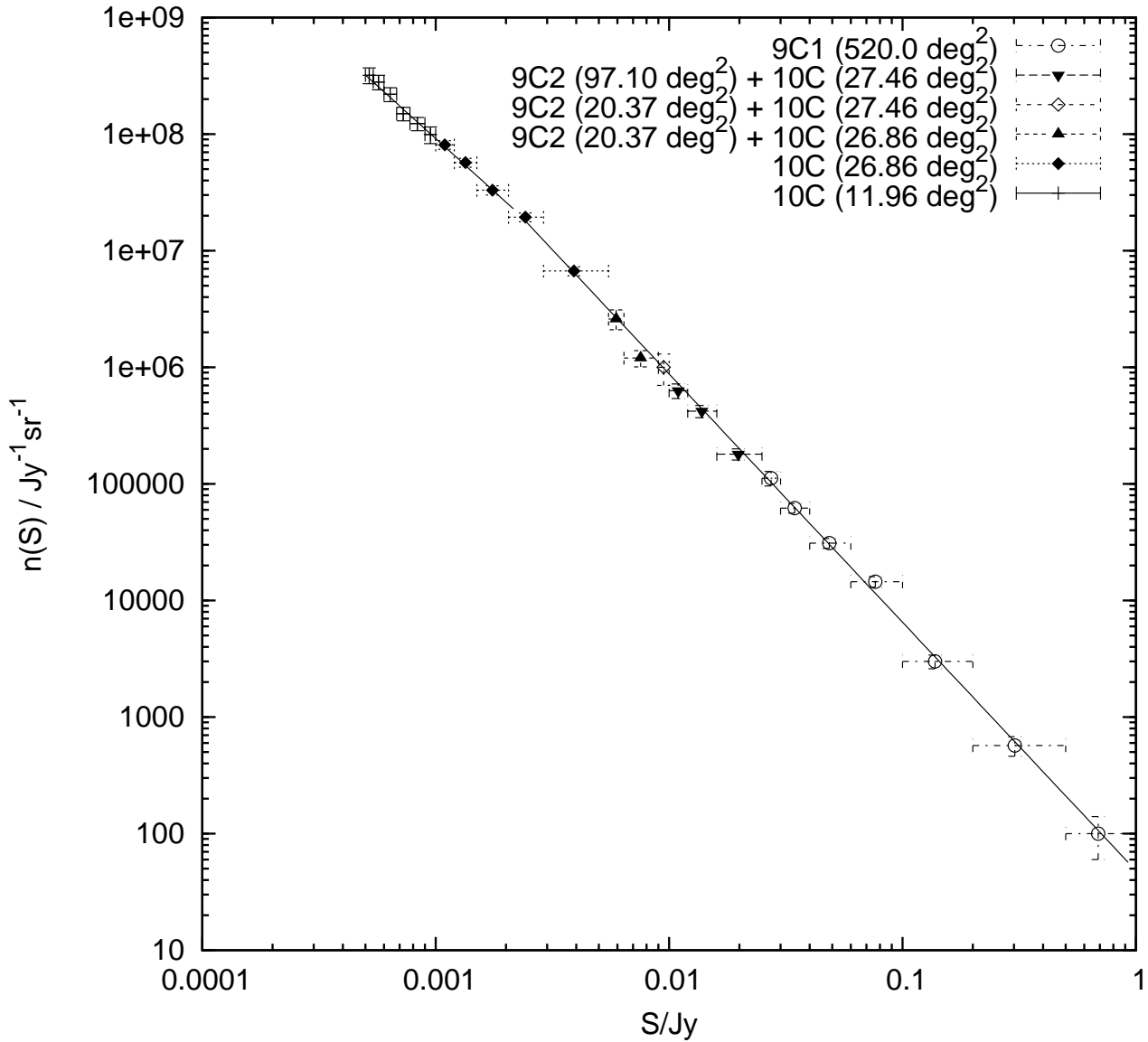}
  \caption{The combined 9C and 10C 15.7-GHz differential source
           count.  Different symbols are used to distinguish between
           the areas that were used to derive the count for the various
           flux-density bins;  `9C1' indicates areas presented in
           \citet{waldram2003}, `9C2' refers to areas presented in
           \citet{waldram2010} and `10C' is used to designate areas
           presented in this paper.  The fitted count is indicated
           by the solid line.}
  \label{fig:joint_count}
 \end{center}
\end{figure*}

Fig.~\ref{fig:joint_count} shows the combined 9C and 10C
differential source count.  Data from areas of the 9C survey
presented in \citet{waldram2003} and \citet{waldram2010} have been
used in constructing the count.  The data used for each bin are
shown in Table~\ref{tab:joint_count}.  For the intermediate
flux-density ranges, for which there are data from both surveys,
some of the 9C survey data were excluded.  This was to avoid
double counting the areas that were surveyed as part of both
the 9C and 10C surveys.

Using the same method as described above, a broken power law
was fitted to the binned differential count.  As previously,
a broken power law was found to be a significantly better fit
to the data than a single power law.  Again, an F-test
indicated that the null hypothesis -- that the data are more
likely to have been drawn from the simpler model -- can be
rejected at $> 99.9$~per~cent confidence.  The best-fit
broken-power-law parameterisation of the source count is
\[n(S) \equiv \frac{\mathrm{d}N}{\mathrm{d}S} \approx \left\{
\begin{array}{l l}
~~48 \left(\frac{S}{\mathrm{Jy}}\right)^{-2.13}~\mathrm{Jy^{-1}~sr^{-1}}~\mathrm{for}~2.2~\mathrm{mJy} \leq S \leq 1~\mathrm{Jy} \\
340 \left(\frac{S}{\mathrm{Jy}}\right)^{-1.81}~\mathrm{Jy^{-1}~sr^{-1}}~\mathrm{for}~0.5 \leq  S < 2.2~\mathrm{mJy.}
\end{array} \right. \]
This fitted count is found to give a good fit to the data, with a
reduced chi-squared value of 0.75.  The probability of obtaining
a reduced chi-squared value greater than 0.75 by chance, given
the number (20) of degrees of freedom, is 78 per cent.  The fit
is indicated in Fig.~\ref{fig:joint_count}.

\subsection{Comparison with the de Zotti model}\label{Comparison to the de Zotti model}

\begin{figure}
 \begin{center}
  \includegraphics[scale=0.46,angle=270]{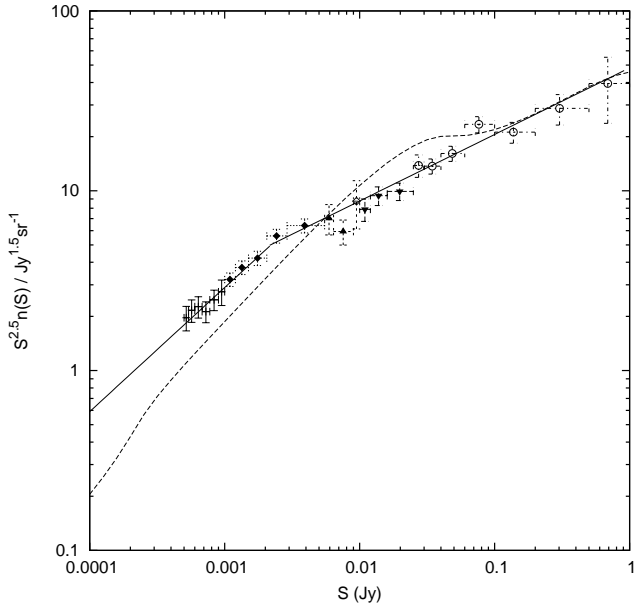}
  \caption{The normalised
           $\left(S^{2.5}\frac{\mathrm{d}N}{\mathrm{d}S}\right)$
           combined 9C and 10C differential source count.  Symbols
           indicating the areas from which the counts were derived
           are identical to those in Fig.~\ref{fig:joint_count}.
           The fitted broken-power-law parameterisation is
           indicated by the solid line.  The dashed line indicates
           the prediction of the latest version of the
           \citet{dezotti2005} model.}
  \label{fig:dezotti_comparison}
 \end{center}
\end{figure}

In Fig.~\ref{fig:dezotti_comparison} the combined 9C and 10C
source count is compared with the latest version of the 15-GHz
source-count model by \citet{dezotti2005}, extracted from their
website\footnote{http://web.oapd.inaf.it/rstools/srccnt/srccnt\_tables}
on 2011 March 01; for completeness, the model counts, over the
relevant flux-density range, are provided in
Appendix~\ref{appendix}.  No attempt has been made to correct for
the small frequency difference between the measured and model
source counts but this is likely to make little difference to the
overall conclusions.

The model count is in good agreement with the measured count at
the high-flux-density end.  However, the shape of the plotted
model count is somewhat different from that of the measured count;
so that, with decreasing flux density, the model first
over-predicts and then, below $\approx 5$~mJy, under-predicts the
measured count.

The total number of sources per steradian with flux densities
between 0.5~mJy and 1~Jy, predicted by the model, was calculated by
integrating the model differential source count between these
limiting flux densities.  Since the predicted counts are given at a
number of discrete flux densities (see
Table~\ref{tab:model_counts}), the integration was carried out
piecewise by approximating the model count as a power law between
successive pairs of points.

The number of sources per unit area predicted by the model was
found to be only 70~per~cent of the measured value.  Because the
differential count is largest at the low flux-density-end, the
under-prediction of the count at the lowest flux densities
dominates over the over-prediction at slightly higher flux
densities, explaining the 30-per-cent deficit over the entire
range.

\section{Matching with 1.4-GH\lowercase{z} surveys}\label{Matching with 1.4-GHz surveys}

The complete and unbiased sample of sources from the deep areas of
the survey (i.e. those sources with flux densities between 0.5 and
25~mJy in these areas) was matched with entries from the 1.4-GHz
catalogues of the Faint Images of the Radio Sky at Twenty
Centimetres \citep[FIRST;][]{becker1995} survey and the
NRAO-VLA Sky Survey \citep[NVSS;][]{condon1998}.  The sample
comprises $\approx 650$ sources in total.  There are NVSS data for
all the 10C survey fields; however, some of the 10C survey fields
are not covered or are only partially covered by FIRST.

The NVSS, which has a resolution of 45~arcsec (slightly lower than
that of the 10C survey) has a completeness of 50~per~cent at
$\approx 2.5$~mJy and 99~per~cent at $\approx 3.4$~mJy.  The
FIRST survey has higher resolution of 5~arcsec and is complete to
1~mJy.  Owing to the significant difference in resolution between
the 10C and FIRST surveys, the flux densities of any FIRST sources
lying within one LA synthesised beam of a 10C source were summed for
comparison with the 10C flux density.

Some results from matching with the NVSS and FIRST catalogues
are shown in Tables~\ref{tab:NVSS_match} and \ref{tab:FIRST_match}
respectively.  In both cases, the sample has been divided into
various 15.7-GHz flux-density bins and the percentage of sources
with flux densities falling in each bin without a corresponding
match in the relevant low-frequency catalogue is indicated.
A limiting 1.4-to-15.7-GHz spectral index, $\alpha_{\mathrm{lim}}$,
based on the completeness
limit of the relevant low-frequency catalogues (3.4~mJy for NVSS and
1.0~mJy for FIRST), is also indicated for each bin.  Any source
with a spectral index greater than this limiting value ought to be
detected in the low-frequency catalogue.

In addition, the percentage of sources with $\alpha > 0.5$ and
$\alpha > 0.81$ is given for each bin.  For the matching to the
NVSS catalogue, for all but the three highest flux-density bins, the
percentages of sources with $\alpha > 0.5$ ought to be regarded
as lower limits.  This is because the values of $\alpha_{\mathrm{lim}}$,
for the lower flux-density bins, are $> 0.5$.
The choice of the second value of 0.81
avoids this problem, since for all flux-density bins and for
matching to both low-frequency catalogues, all sources with
$\alpha > 0.81$ should be matched.

It is noted that the percentages of sources with $\alpha > 0.5$
and $\alpha > 0.81$ from matching to the FIRST survey are
systematically lower than those from matching to the NVSS.  This
is most likely due to the fact that extended emission detected
as part of the NVSS is resolved out as part of the FIRST survey.

An interesting effect, whereby the source population shifts
towards a flatter-spectrum population with decreasing flux density,
is observed in the data.  This is most clearly seen in
the matching to the NVSS from the change in the fraction of sources
with $\alpha > 0.81$.  As mentioned previously, the fraction of
sources with $\alpha > 0.5$ represents a lower limit for a number
of bins and so the evolution in this fraction cannot readily be
interpreted as evidence for a change in the population.  In the
matching to the FIRST survey, the effect is best demonstrated by
the change in the fraction of sources with $\alpha > 0.5$.  Owing
to the fact that FIRST coverage is not available for all the 10C
fields, the statistics for the fractions of sources with
$\alpha > 0.81$ are poorer compared with those for the matching with
the NVSS.  Also, the interpretation of the results is complicated
by the mismatch in resolution noted above.

\begin{table*}
\begin{center}
\caption{Some statistics from matching with the NVSS source catalogue.
         Numbers and percentages of sources with $\alpha > 0.5$ for the
         lowest four flux-density bins are lower limits.}
\label{tab:NVSS_match}
\begin{tabular}{@{} c d d c d d d d d c c}
\hline
Bin start&\dhead{Bin end}  &\ehead{Total}&Matched&\dhead{Unmatched}&\dhead{$\alpha_{\mathrm{lim}}$}&\dhead{Percentage}&\alpha > 0.5&\ehead{Percentage}&$\alpha > 0.81$&Percentage \\
$S$ (mJy)&\dhead{$S$ (mJy)}&             &       &                 &                               &\dhead{Unmatched} &            &\alpha > 0.5      &        & $\alpha > 0.81$  \\
\hline
5.0      & 25.0            & 55         & 54     & 1               & -0.15                         & 2                &   38       &   69             & 27            & 49        \\
2.5      & 5.0             & 65         & 62     & 3  	           &  0.14                         & 5                &   47       &   72             & 30            & 46        \\
1.5      & 2.5             & 99         & 87     & 12 	           &  0.35                         & 12               &   64       &   74             & 45            & 46        \\
1.0      & 1.5             & 114        & 76     & 38 	           &  0.52                         & 33               & > 67       & > 59             & 29            & 25        \\
0.8      & 1.0             & 79         & 50     & 29 	           &  0.61                         & 37               & > 42       & > 53             & 22            & 28        \\
0.6      & 0.8             & 125        & 60     & 65 	           &  0.73                         & 52               & > 59       & > 47             & 26            & 21        \\
0.5      & 0.6             & 107        & 43     & 64 	           &  0.81                         & 60               & > 43       & > 40             & 19            & 18        \\
\hline
\end{tabular}
\end{center}
\end{table*}

\begin{table*}
\begin{center}
\caption{Some statistics from matching with the FIRST source catalogue.}
\label{tab:FIRST_match}
\begin{tabular}{@{}c d c c d d d c c d c }
\hline
Bin start&\dhead{Bin end}  &Total&Matched& \dhead{Unmatched} & \dhead{$\alpha_{\mathrm{lim}}$} & \dhead{Percentage} & $\alpha > 0.5$ & Percentage     & \alpha > 0.81 & Percentage\\
$S$ (mJy)&\dhead{$S$ (mJy)}&     &       &                   &                                 & \dhead{Unmatched}  &                & $\alpha > 0.5$ &         & $\alpha > 0.81$ \\
\hline
5.0      & 25.0            & 36  & 36    & 0                 & -0.67                           & 0                  & 17             & 47  	      & 8 	      & 22        \\
2.5      & 5.0             & 46  & 44    & 2  	             & -0.38                           & 4                  & 23	     & 50  	      & 7 	      & 15        \\
1.5      & 2.5             & 56  & 50    & 6  	             & -0.17                           & 11                 & 30	     & 54  	      & 8 	      & 14        \\
1.0      & 1.5             & 67  & 50    & 17 	             &  0.00                           & 25                 & 24	     & 36  	      & 10            & 15        \\
0.8      & 1.0             & 50  & 41    & 9  	             &  0.09                           & 18                 & 18	     & 36  	      & 5 	      & 10        \\
0.6      & 0.8             & 88  & 56    & 32 	             &  0.21                           & 36                 & 35	     & 40  	      & 15	      & 17        \\
0.5      & 0.6             & 68  & 32    & 36 	             &  0.29                           & 53                 & 18	     & 27  	      & 7 	      & 10        \\
\hline
\end{tabular}
\end{center}
\end{table*}

By matching sources detected as part of the 9C survey to the
NVSS catalogue, \citet{waldram2010} observed a shift in the
15-GHz-band source population from being dominated by flat-spectrum
to being dominated by steep-spectrum sources with decreasing flux
density between $\approx 100$~mJy and $\approx 10$~mJy -- this
effect is illustrated by Fig.~\ref{fig:sindex_correction}.  A
similar effect has been observed by \citet{massardi2010} by
matching entries from their 20-GHz source catalogue to
low-frequency catalogues.  Their large sample size allows them to
trace the spectral evolution of the source population in detail
between 40~mJy and 1~Jy.  The median value of $\alpha^{20}_{1.4}$ is
found to become rapidly larger with decreasing flux density below
$\approx 80$~mJy (E.~Sadler; private communication).

The 10C data suggest that, at fainter flux densities, this trend is
reversed with a move back towards a flatter-spectrum population.  A
shift towards a flatter-spectrum population has been observed in
sub-mJy data from low-frequency surveys
\citep[see, for example,][and references therein]{prandoni2006}.

\section{Extended sources}\label{Extended sources}

Using the methods described in Paper~I, 4.8~per~cent of the 10C
sources are classified as extended and 5.9~per~cent as overlapping
(the categories are mutually exclusive).  Some of the `extended'
sources might be classified as such owing to source confusion or
effects arising from noise.  Similarly, chance juxtapositions of
sources are likely to result in sources with no genuine association
being classified as overlapping.  Consequently, simulations in
which simulated point sources were inserted, at randomly-selected
positions, into the 10C survey fields were carried out in an
attempt to provide an estimate of the fraction of 10C sources that
are genuinely extended relative to the LA synthesised beam.

In total $\approx 1600$ simulated point sources were added into the
10C data in the \textit{uv} plane.  The sources were simulated with
flux densities drawn randomly from the measured source count.  As
for the simulation to assess the survey completeness, no simulated
source was allowed to lie within 2~arcmin of any other.  The
standard mapping and source extraction procedures, described in
Paper~I, were applied to the data.  Of the simulated point sources
that were detected, $1.4 \pm 0.5$~per~cent were classified as
extended and $2.2 \pm 0.4$~per~cent were categorised as
overlapping, where the uncertainties have been estimated from the
scatter in the results between the different survey fields.

The results from the simulation suggest that
$\approx 100 \times (4.8 - 1.4)/4.8 \approx 71$~per~cent of the
sources classified as such are genuinely extended and that
$\approx 100 \times (5.9 - 2.2)/5.9 \approx 63$~per~cent of
overlapping sources are actually a component of a larger object.
In the great majority of cases, overlapping sources are found to
have just two components.  Therefore, if the two categories are
taken together -- that is, sources consisting of a single component
that is extended relative to the telescope synthesised beam and
extended sources consisting of several overlapping components --
the fraction of 10C sources that are genuinely extended relative to
the LA synthesised beam of $\approx 30$~arcsec is estimated as
$(5.9 - 2.2)/2~+~(4.8 - 1.4) \approx 5$~per~cent (the factor of
$1/2$ results from the assumption that the overlapping sources
consist of two components).  This result assumes no clustering of
sources on small ($\lesssim 1$~arcmin) angular scales.  It is
noted that a similar fraction ($\approx 6$ per cent) of extended
sources was measured from 9C data by \citet{waldram2003}, who
carried out the classification on the basis of manual inspection
of 9C contour maps.

Contour plots of all the sources identified by the automated
source-extraction procedures as extended or overlapping were
inspected by eye.  Sources thought most likely to be genuinely
extended from their morphologies (for example overlapping sources
with similar peak flux densities) were identified.  10C contour
plots of 39 such sources were compared with contour plots from the
NVSS and, where available, from the FIRST survey.  In addition, the
NASA Extragalactic Database\footnote{http://nedwww.ipac.caltech.edu}
(NED) was searched for objects detected within $\approx 2$~arcmin
of each of the extended 10C sources.  NED reports entries from
catalogues belonging to a number of low-frequency radio surveys, in
addition to entries from the catalogues of optical, infrared and
X-ray surveys.

From these investigations 29 of the 39 sources were found to be
genuinely extended and two artefacts of source confusion.  In eight
cases it still remained unclear whether the sources were genuinely
extended.  Most of the genuinely extended sources were found to be
classical doubles but three are nearby galaxies and four are
twin-jet/tailed sources.  One of the twin-jet sources has
$z = 0.267$ and another has $z = 0.391$; the respective centroid
positions of these sources are 15:41:05.0 +43:27:00 and 10:44:56.7
+59:25:38.  Some results for one source of particular interest are
provided below.

\subsection{8C~0821+695}

\begin{figure}
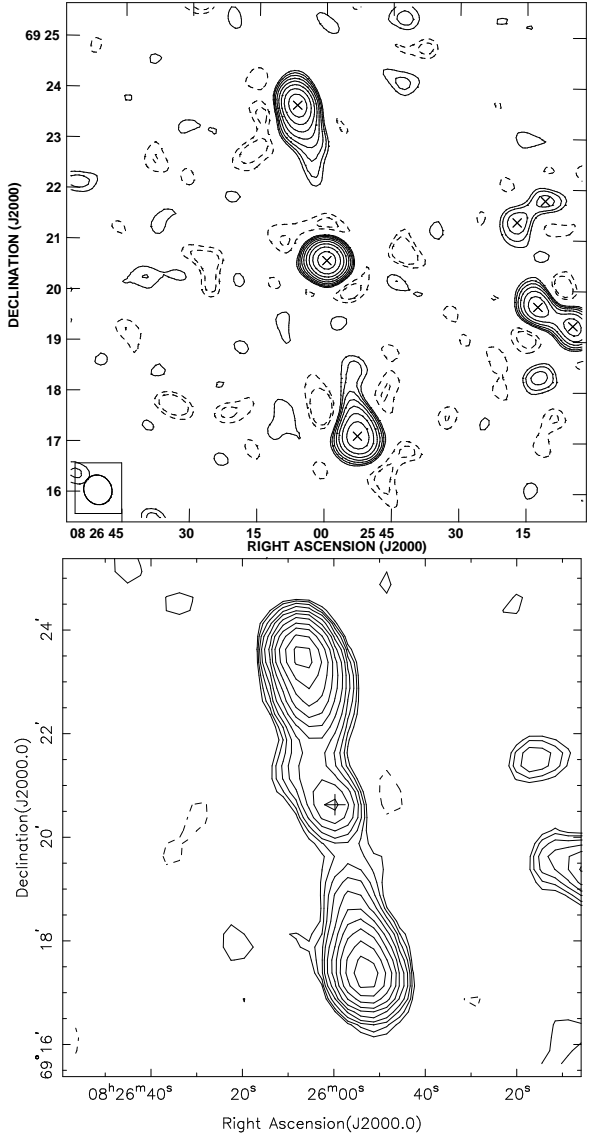

 \begin{center}
  \includegraphics[scale=0.40,angle=270]{10CJ082559+692035.epsi}
  \includegraphics[scale=0.435,angle=270,clip=]{10CJ082559+692037_NVSS.epsi}
  \caption{Contour plots of 8C~0821+695, a giant radio galaxy lying
           in one of the 10C survey fields, from the 10C survey at
           15.7~GHz (top) and the NVSS at 1.4~GHz.  In both cases
           a factor of $\sqrt{2}$ separates the contours, which
           start at $\pm 0.1$ and $\pm 1.0$~mJy~beam$^{-1}$
           for the 10C and NVSS plots respectively.  The negative
           contours are shown using dashed lines.  Crosses in the
           10C plot indicate the position of 10C sources.  The cross
           in the NVSS plot gives the position of the central
           component source at 15.7~GHz.}
  \label{fig:linear_triple}
 \end{center}
\end{figure}

It was noticed, whilst manually inspecting the contour plots of
two sources flagged as extended, that these sources were the lobes
of a `linear triple', a broad-emission-line radio galaxy or
radio-loud quasar (see Fig.~\ref{fig:linear_triple}).  The central
component and outer lobes span an angular distance of 6.6~arcmin and
appear as three separate sources in the 10C catalogue, since no
linking emission is detected using the 10C data.  The source,
8C~0821+695, was noticed in the 8C survey data \citep{rees1990}
by \citet{lacy1993}, who measured the size of the
source to be 1.5~h$^{-1}$~Mpc and to have a redshift of 0.538.  They
estimate the age of the source to be $\sim 3 \times 10^{7}$~yr and
argue that the source is large because it has expanded rapidly in a
low-density environment rather than because it is old.  Assuming no
variability, the 10C data imply spectral steepening of the emission
from the central component; \citet{lacy1993}
found a spectral index for the central component of $0.23 \pm 0.04$ between 1.4 and 5~GHz.  The
10C measurements imply a spectral index for the central component of $0.79 \pm 0.06$ between 5
and 15.7~GHz.

\section{Conclusions}\label{Conclusions}

The AMI LA has been used to carry out the 10C survey, the deepest
radio source survey of any significant extent
($\gtrsim 0.2$~deg$^{2}$) above 1.4~GHz.  The resulting deep
15.7-GHz source counts are useful for the interpretation of CMB
data, for which foreground radio sources are an important
contaminant.  The source catalogue also provides an invaluable
resource for the study of faint high-frequency-selected radio
sources.  The survey covers $\approx 27$~deg$^{2}$ complete to
1~mJy and $\approx 12$~deg$^2$ (wholly contained within the larger
area) complete to 0.5~mJy.  The number of sources with
$S > 25$~mJy, appearing in the survey catalogue, is biased low;
several of the survey fields were chosen to minimise the number of
such sources appearing within the survey areas.  In total, 1897
sources appear in the 4.62-$\sigma$ catalogue; the faintest being
$\approx 100$~$\mu$Jy.  A number of the key conclusions from the
work are listed below.

\begin{noindlist}
\item[(1)] The 10C differential source count was parameterised
using a broken power law.  This was found to provide a
significantly better fit to the data than a simpler,
single-power-law parameterisation.  The best-fit broken power law
is
\[n(S) \equiv \frac{\mathrm{d}N}{\mathrm{d}S} \approx \left\{
\begin{array}{l l}
~~24 \left(\frac{S}{\mathrm{Jy}}\right)^{-2.27}~\mathrm{Jy^{-1}~sr^{-1}}~\mathrm{for}~2.8 \leq S \leq 25~\mathrm{mJy} \\
376 \left(\frac{S}{\mathrm{Jy}}\right)^{-1.80}~\mathrm{Jy^{-1}~sr^{-1}}~\mathrm{for}~0.5 \leq  S < 2.8~\mathrm{mJy.}
\end{array} \right. \]
\newline

\item[(2)] After having applied corrections to the individual
source flux densities measured as part of the 9C survey to account
for the small difference in frequencies between the 9C and 10C
surveys, the 9C and 10C data were combined.  The addition of the 9C
data allowed the calculation of the best estimate of the 15.7-GHz
differential source count by improving the source-count statistics
for $5.5 \leq S < 25$~mJy and by providing data for a complete
sample of sources with $25~\mathrm{mJy} \leq S \leq 1~\mathrm{Jy}$.
Again, a broken-power-law parameterisation was found to offer a
significantly improved fit to the data compared with that provided by
a single power law.  The best-fit broken power law is
\[n(S) \equiv \frac{\mathrm{d}N}{\mathrm{d}S} \approx \left\{
\begin{array}{l l}
~~48 \left(\frac{S}{\mathrm{Jy}}\right)^{-2.13}~\mathrm{Jy^{-1}~sr^{-1}}~\mathrm{for}~2.2~\mathrm{mJy} \leq S \leq 1~\mathrm{Jy} \\
340 \left(\frac{S}{\mathrm{Jy}}\right)^{-1.81}~\mathrm{Jy^{-1}~sr^{-1}}~\mathrm{for}~0.5 \leq  S < 2.2~\mathrm{mJy.}
\end{array} \right. \]
\newline

\item[(3)] The model counts by \citet{dezotti2005} are found to
display good agreement with the 9C and 10C data at the high
flux-density-end of the measured count.  However, with decreasing
flux density the model first over-predicts and then, below about
5~mJy, under-predicts the measured count.  By integrating the model
differential source count, the model was found to under-predict the
total number of sources, with flux densities between 0.5~mJy and
1~Jy, per unit area by approximately 30~per~cent.  This deficit,
over the entire flux-density range, is attributable to the
model underestimating the count at the lowest flux densities.
\newline

\item[(4)] Entries from the 10C source catalogue were matched to
those contained in the catalogues of the NVSS and FIRST survey
(both of which have observing frequencies of 1.4~GHz).  The
matching revealed a shift in the typical 1.4-to-15.7-GHz spectral
index of the 15.7-GHz-selected source population with decreasing
flux density towards sub-mJy levels.  When matching to NVSS, 49
per cent of sources with $5.0 \leq S_{10C} < 25.0$~mJy were found
to have $\alpha^{15.7}_{1.4} > 0.81$.  However, for sources with
$0.5 \leq S_{10C} < 0.6$~mJy the corresponding figure is 18 per
cent.  The observed trend is in contrast to that measured for
sources with higher flux densities by \citet{waldram2010}.  They
found that the typical spectral index became steeper for sources
with  decreasing flux densities between $\approx 100$~mJy and
$\approx 10$~mJy.  A similar effect, to that measured using the
10C data, has been observed as part of lower-frequency surveys --
in lower-frequency-selected samples significant numbers of flatter
spectrum sources start to enter at sub-mJy levels
\citep[see, for example,][and references therein]{prandoni2006}.
\newline

\item[(5)]  Automated techniques for identifying extended sources,
described in Paper~I, have been applied to the data.  The
proportion of sources that are extended relative to the LA
synthesised beam of $\approx 30$~arcsec is $\approx 5$~per~cent;
this is similar to the proportion of $\approx 6$ per cent measured
by \citet{waldram2003}.  A subset of 39 extended or overlapping
sources, thought likely to be genuinely extended on the basis of
their 15.7-GHz morphologies, were investigated further using
higher-resolution data.  These data confirmed that at least 29 of
these sources are genuinely extended; most were identified as
classical doubles but three are nearby galaxies and four are
twin-jet sources.
\end{noindlist}

\section*{Acknowledgments}\label{acknowledgements}

We would like to thank the staff of the Mullard Radio Astronomy
Observatory for maintaining and operating AMI.  We are also
grateful to the University of Cambridge and PPARC/STFC for funding
and supporting the LA.  MLD, TMOF, MO, CRG, MPS and TWS acknowledge
support from PPARC/STFC studentships.  We would also like to thank
Elaine Sadler and Gianfranco de~Zotti for their helpful comments,
which have improved this paper, and Sally Hales for managing the
online catalogues.


\appendix\section{Model source counts}\label{appendix}

Table~\ref{tab:model_counts} shows the 15-GHz model source counts
by \citet{dezotti2005} to which the measured 9C and 10C source
counts are compared in
Section~\ref{Comparison to the de Zotti model}.  Here, the model
counts are provided only for the flux-density range relevant for
comparison to 9C and 10C data.

\begin{table}
 \begin{center}
  \caption{15-GHz model source counts by \citet{dezotti2005},
           presented as logarithmic, normalised differential
           counts -- specifically
           $\log_{10}(S^{5/2} \mathrm{d}N/\mathrm{d}S)$.
           Predicted counts are given for a number
           of different source populations; the flat-spectrum
           counts are made up from the sum of the
           flat-spectrum-radio-quasar (FSRQ) and BL-Lacertae
           (BL Lac) counts, and the total counts are found by
           summing the flat-spectrum and steep-spectrum counts.}
  \label{tab:model_counts}
  \begin{tabular}{@{} d d d d d d}
  \hline
  \dhead{$\log_{10} S$} & \multicolumn{5}{c}{Source count (Jy$^{1.5}$~sr$^{-1}$)} \\
  \dhead{(Jy)}          & \dhead{FSRQ} & \dhead{BL Lac}   & \dhead{Flat}     & \dhead{Steep}    & \dhead{Total}   \\
                        &              &                  & \dhead{spectrum} & \dhead{spectrum} &                 \\
  \hline      						          
  -4.000                & -1.563       & -1.020           & -0.911           & -1.084           & -0.688          \\
  -3.900                & -1.479       & -0.972           & -0.854           & -0.935           & -0.592          \\
  -3.800                & -1.395       & -0.924           & -0.797           & -0.778           & -0.487          \\
  -3.700                & -1.311       & -0.876           & -0.740           & -0.609           & -0.369          \\
  -3.600                & -1.226       & -0.829           & -0.682           & -0.445           & -0.247          \\
  -3.500                & -1.142       & -0.781           & -0.624           & -0.319           & -0.144          \\
  -3.400                & -1.058       & -0.735           & -0.566           & -0.215           & -0.055          \\
  -3.300                & -0.974       & -0.688           & -0.507           & -0.119           &  0.030          \\
  -3.200                & -0.890       & -0.642           & -0.447           & -0.028           &  0.112          \\
  -3.100                & -0.806       & -0.596           & -0.387           &  0.059           &  0.192          \\
  -3.000                & -0.721       & -0.550           & -0.326           &  0.146           &  0.272          \\
  -2.900                & -0.637       & -0.505           & -0.265           &  0.232           &  0.352          \\
  -2.800                & -0.553       & -0.460           & -0.203           &  0.316           &  0.431          \\
  -2.700                & -0.469       & -0.415           & -0.140           &  0.400           &  0.510          \\
  -2.600                & -0.385       & -0.371           & -0.077           &  0.483           &  0.588          \\
  -2.500                & -0.301       & -0.327           & -0.013           &  0.564           &  0.666          \\
  -2.400                & -0.217       & -0.283           &  0.052           &  0.643           &  0.742          \\
  -2.300                & -0.133       & -0.240           &  0.118           &  0.721           &  0.818          \\
  -2.200                & -0.050       & -0.197           &  0.184           &  0.795           &  0.890          \\
  -2.100                &  0.034       & -0.155           &  0.251           &  0.868           &  0.962          \\
  -2.000                &  0.118       & -0.113           &  0.319           &  0.935           &  1.029          \\
  -1.900                &  0.201       & -0.071           &  0.387           &  0.998           &  1.093          \\
  -1.800                &  0.285       & -0.029           &  0.456           &  1.056           &  1.153          \\
  -1.700                &  0.368       &  0.012           &  0.526           &  1.103           &  1.205          \\
  -1.600                &  0.451       &  0.052           &  0.597           &  1.140           &  1.249          \\
  -1.500                &  0.534       &  0.092           &  0.668           &  1.163           &  1.284          \\
  -1.400                &  0.616       &  0.132           &  0.739           &  1.163           &  1.302          \\
  -1.300                &  0.698       &  0.172           &  0.811           &  1.137           &  1.305          \\
  -1.200                &  0.779       &  0.211           &  0.883           &  1.103           &  1.308          \\
  -1.100                &  0.860       &  0.250           &  0.955           &  1.076           &  1.321          \\
  -1.000                &  0.939       &  0.288           &  1.027           &  1.052           &  1.340          \\
  -0.900                &  1.018       &  0.326           &  1.098           &  1.028           &  1.366          \\
  -0.800                &  1.095       &  0.363           &  1.169           &  1.003           &  1.395          \\
  -0.700                &  1.170       &  0.401           &  1.238           &  0.977           &  1.428          \\
  -0.600                &  1.242       &  0.438           &  1.305           &  0.951           &  1.464          \\
  -0.500                &  1.310       &  0.474           &  1.369           &  0.924           &  1.502          \\
  -0.400                &  1.373       &  0.510           &  1.429           &  0.898           &  1.541          \\
  -0.300                &  1.430       &  0.546           &  1.483           &  0.873           &  1.579          \\
  -0.200                &  1.479       &  0.581           &  1.531           &  0.850           &  1.613          \\
  -0.100                &  1.517       &  0.616           &  1.568           &  0.827           &  1.641          \\
   0.000                &  1.543       &  0.651           &  1.595           &  0.805           &  1.660          \\
  \hline
  \end{tabular}
 \end{center}
\end{table}
\label{lastpage}
\end{document}


\setcounter{page}{16}

\begin{table*}
\setcounter{table}{0}
\centering
\begin{minipage}{255mm}
\caption{Samples from the 10C source catalogue, which is available in full online at www.mrao.cam.ac.uk/surveys/10C.
         The extraction procedures for the various parameters are described fully in Paper~I.  The columns are the
         source name; the group name, which indicates overlapping sources (the group is named after the source with
         the highest peak flux density in the group and the number of objects in the group is given in brackets);
         the right ascension, $\alpha_{\mathrm{pk}}$, and declination, $\delta_{\mathrm{pk}}$, (J2000) of the source
         peak; the peak flux density, $S_{\mathrm{pk}}$, and its uncertainty, $\delta S_{\mathrm{pk}}$; the right
         ascension, $\alpha_{\mathrm{in}}$, and declination, $\delta_{\mathrm{in}}$, of the source centroid (J2000;
         relevant for extended sources); the integrated flux density, $S_{\mathrm{in}}$, and its uncertainty,
         $\delta S_{\mathrm{in}}$; the critical value, $e_{\mathrm{crit}}$, of the major axis, above which the
         source is defined as extended, having deconvolved the telescope's point-source response from the fitted 2-D
         Gaussian; the lengths, $e_{\mathrm{maj}}$ and $e_{\mathrm{min}}$, of the major and minor axes and position
         angle, $e_{\theta}$, of the source after the deconvolution -- the $e_{\mathrm{min}}$ and $e_{\theta}$
         columns are left blank for sources classified as point-like; the source type, $t$ -- point-like (P) or
         extended (E); a flag, `*', indicating that the error from the approximation of the point-source response
         is significant and, consequently, the results from the Gaussian fitting ought to be treated with caution;
         and the region of the survey in which the source falls -- deep (D), shallow (S) or neither (N).  Note
         that the parameters from the Gaussian fitting are omitted if the fitting did not converge.}
\label{table:sample_catalogue}
\begin{tabular}{@{} c c c c r r c c r r r r r r c c c }
\hline
Source name
&Group name
&$\alpha_{\mathrm{pk}}$
&$\delta_{\mathrm{pk}}$
&\multicolumn{1}{c}{$S_{\mathrm{pk}}$}
&\multicolumn{1}{c}{$\delta S_{\mathrm{pk}}$}
&\multicolumn{1}{c}{$\alpha_{\mathrm{in}}$}
&\multicolumn{1}{c}{$\delta_{\mathrm{in}}$}
&\multicolumn{1}{c}{$S_{\mathrm{in}}$}
&\multicolumn{1}{c}{$\delta S_{\mathrm{in}}$}
&\multicolumn{1}{c}{$e_{\mathrm{crit}}$}
&\multicolumn{1}{c}{$e_{\mathrm{maj}}$}
&\multicolumn{1}{c}{$e_{\mathrm{min}}$}
&\multicolumn{1}{c}{$e_{\theta}$}
&$t$&F&R\\
&&&&\multicolumn{1}{c}{(mJy)}&\multicolumn{1}{c}{(mJy)}&&&\multicolumn{1}{c}{(mJy)}&\multicolumn{1}{c}{(mJy)}&\multicolumn{1}{c}{($''$)}&\multicolumn{1}{c}{($''$)}&\multicolumn{1}{c}{($''$)}&\multicolumn{1}{c}{($^{\circ}$)}&&l&e\\
&&&&&&&&&&&&&&&a&g\\
&&&&&&&&&&&&&&&g&i\\
&&&&&&&&&&&&&&& &o\\
&&&&&&&&&&&&&&& &n\\
\hline
10CJ002129+322700 &                      & 00:21:29.8 & +32:27:00 & 24.69 & 1.34 & 00:21:29.9 & +32:27:01 & 23.27 & 1.28 & 25.0 & 13.5 &     &       & P & * & S \\
                  &                      &            &           &       &      &            &           &       &      &      &      &     &       &   &   &   \\
10CJ002215+314923 &                      & 00:22:15.8 & +31:49:23 &  0.51 & 0.07 & 00:22:15.8 & +31:49:25 &  0.40 & 0.09 & 43.2 &  0.0 &     &       & P &   & D \\
10CJ002215+321707 &                      & 00:22:15.8 & +32:17:07 &  1.55 & 0.10 & 00:22:15.8 & +32:17:07 &  1.47 & 0.13 & 25.0 &  9.5 &     &       & P &   & D \\
10CJ002216+315153 &                      & 00:22:16.2 & +31:51:53 &  0.64 & 0.07 & 00:22:16.2 & +31:51:52 &  0.56 & 0.10 & 39.1 &  0.0 &     &       & P &   & D \\
10CJ002219+314717 &                      & 00:22:19.3 & +31:47:17 &  0.83 & 0.08 & 00:22:19.3 & +31:47:19 &  0.76 & 0.11 & 34.2 & 11.9 &     &       & P &   & D \\
10CJ002220+323537 & 10CJ002222+323549(2) & 00:22:20.0 & +32:35:37 &  2.15 & 0.20 & 00:22:19.4 & +32:35:31 &  3.51 & 0.44 & 32.5 & 58.2 & 0.0 &  49.7 & E &   & S \\
10CJ002221+314422 &                      & 00:22:21.9 & +31:44:22 &  0.67 & 0.07 & 00:22:22.0 & +31:44:21 &  0.61 & 0.10 & 36.3 &  0.0 &     &       & P &   & D \\
10CJ002222+322803 &                      & 00:22:22.1 & +32:28:03 &  0.77 & 0.15 & 00:22:22.0 & +32:28:04 &  0.77 & 0.24 & 52.3 &  9.4 &     &       & P &   & S \\
10CJ002222+323549 & 10CJ002222+323549(2) & 00:22:22.3 & +32:35:49 &  2.40 & 0.21 & 00:22:22.8 & +32:35:51 &  3.17 & 0.40 & 30.9 & 44.9 & 0.0 &  41.7 & E &   & S \\
10CJ002223+312509 & 10CJ002224+312409(3) & 00:22:23.3 & +31:25:09 &  1.46 & 0.10 & 00:22:23.4 & +31:25:10 &  1.36 & 0.12 & 25.0 &  5.3 &     &       & P &   & D \\
10CJ002224+313627 &                      & 00:22:24.4 & +31:36:27 &  0.62 & 0.06 & 00:22:24.4 & +31:36:29 &  0.74 & 0.11 & 35.5 & 31.9 &     &       & P &   & D \\
10CJ002224+312409 & 10CJ002224+312409(3) & 00:22:24.9 & +31:24:09 &  6.86 & 0.37 & 00:22:24.9 & +31:24:10 &  7.14 & 0.40 & 25.0 &  9.1 &     &       & P &   & D \\
10CJ002227+312520 & 10CJ002224+312409(3) & 00:22:27.4 & +31:25:20 &  0.66 & 0.07 & 00:22:27.3 & +31:25:19 &  0.59 & 0.10 & 36.5 &  4.3 &     &       & P &   & D \\
10CJ002232+315213 &                      & 00:22:32.1 & +31:52:13 &  0.44 & 0.07 & 00:22:32.1 & +31:52:15 &  0.56 & 0.13 & 47.8 & 33.7 &     &       & P &   & D \\
10CJ002233+323802 &                      & 00:22:33.4 & +32:38:02 &  1.04 & 0.18 & 00:22:33.6 & +32:38:02 &  0.97 & 0.27 & 47.9 &  7.6 &     &       & P &   & S \\
10CJ002237+310910 &                      & 00:22:37.7 & +31:09:10 &  1.50 & 0.17 & 00:22:37.6 & +31:09:11 &  1.51 & 0.27 & 40.7 & 17.2 &     &       & P &   & S \\
10CJ002238+312622 &                      & 00:22:38.6 & +31:26:22 &  0.88 & 0.08 & 00:22:38.7 & +31:26:23 &  0.84 & 0.11 & 31.0 &  7.1 &     &       & P &   & D \\
10CJ002240+314503 &                      & 00:22:40.1 & +31:45:03 &  0.41 & 0.07 & 00:22:40.0 & +31:45:00 &  0.51 & 0.13 & 47.9 & 44.9 &     &       & P &   & D \\
10CJ002241+310007 &                      & 00:22:41.3 & +31:00:07 &  1.34 & 0.20 & 00:22:41.3 & +31:00:06 &  1.33 & 0.33 & 49.7 & 16.7 &     &       & P &   & S \\
10CJ002241+311704 &                      & 00:22:41.3 & +31:17:04 &  0.89 & 0.14 & 00:22:41.2 & +31:17:04 &  0.92 & 0.23 & 49.2 & 15.9 &     &       & P &   & S \\
10CJ002242+311844 &                      & 00:22:42.3 & +31:18:44 &  0.84 & 0.11 & 00:22:42.2 & +31:18:43 &  1.28 & 0.24 & 40.9 & 43.7 & 0.0 &   0.2 & E &   & S \\
10CJ002246+313113 &                      & 00:22:46.0 & +31:31:13 &  3.33 & 0.19 & 00:22:45.8 & +31:31:16 &  5.22 & 0.31 & 25.0 & 42.9 & 9.0 & 145.3 & E &   & D \\
10CJ002246+320614 &                      & 00:22:46.1 & +32:06:14 &  0.47 & 0.07 & 00:22:46.2 & +32:06:15 &  0.40 & 0.10 & 43.8 & 11.6 &     &       & P &   & D \\
10CJ002249+310336 &                      & 00:22:49.3 & +31:03:36 &  2.82 & 0.21 & 00:22:49.4 & +31:03:37 &  2.96 & 0.30 & 29.4 & 10.9 &     &       & P &   & S \\
10CJ002251+322305 &                      & 00:22:51.3 & +32:23:05 &  0.50 & 0.08 & 00:22:51.3 & +32:23:06 &  0.39 & 0.10 & 45.7 &  0.0 &     &       & P &   & D \\
10CJ002252+320204 &                      & 00:22:52.7 & +32:02:04 &  0.46 & 0.05 & 00:22:52.7 & +32:02:06 &  0.52 & 0.09 & 37.1 & 27.6 &     &       & P &   & D \\
\hline
\end{tabular}
\end{minipage}
\label{parameters}
\end{table*}